\documentclass[a4paper,11.5pt]{article}
\usepackage{notations}
\usepackage[utf8]{inputenc}
\usepackage[T1]{fontenc}
\usepackage{lmodern}
\usepackage[margin=1.0in]{geometry}
\newtheorem{theorem}{Theorem}[section]
\newtheorem{lemma}[theorem]{Lemma}
\newtheorem{defn}[theorem]{Definition}
\usepackage[numbers]{natbib}
\usepackage{microtype}
\begin{document}

\author[1]{Ahmed Ali Abbasi}
\author[2]{Shuchin Aeron}
\author[3]{Abiy Tasissa \thanks{Corresponding Author: abiy.tasissa@Tufts.edu}}
\affil[1]{\small Department of Electical and Computer Engineering, Iowa State University, Ames, Iowa 50011, USA.}
\affil[2]{\small Department of Electrical and Computer Engineering, Tufts University, Medford, MA 02155,
USA.}
\affil[3]{\small Department of Mathematics, Tufts University, Medford, MA 02155, USA.}
\title{Alternating Minimization algorithm for unlabeled sensing and linked linear regression}
\maketitle

\begin{abstract}
Unlabeled sensing  is a linear inverse problem with permuted  measurements. 
We propose an alternating minimization (AltMin) algorithm with a suitable initialization for two widely considered permutation models:  partially shuffled/$k$-sparse permutations  and $r$-local/block diagonal permutations. Key to the performance of the AltMin algorithm is the  initialization. For the exact unlabeled sensing problem, assuming either a Gaussian measurement matrix or a sub-Gaussian signal, we  bound the initialization error in terms of the number of blocks $s$ and the number of shuffles $k$.  
Experimental results show that our algorithm is fast, applicable to both permutation models, and robust to choice of measurement matrix. We also test our algorithm on several real datasets for the `linked linear regression' problem and show superior performance compared to baseline methods. 
\end{abstract}

\maketitle

\section{Introduction}
\label{sec:intro}
The linear inverse problem is given by $\Y = \B \X^* + \W$, where $\B \in \mathbb{R}^{n \times d}$ is the measurement matrix, $\Y \in \mathbb{R}^{n \times m}$ represents the linear measurements of the unknown $\X^* \in \mathbb{R}^{d \times m}$, and $\W_{ij} \sim \mathcal{N}(0, \sigma^2)$ denotes i.i.d. Gaussian noise. In unlabeled sensing, the measurements $\Y$ are scrambled. Specifically, 
\begin{equation}
\Y = \P^* \B \X^* + \W, \label{eq:model}
\end{equation}
where $\P^* \in \mathbb{R}^{n \times n}$ is the unknown permutation matrix.  Given $\Y$ and $\B$, the objective is to estimate $\X^*$. In this manuscript, $\X^*$ and $\P^*$ denote the underlying unknown parameters of the unlabeled sensing problem.  If there is no noise, i.e., $\mathbf{W}=\mathbf{0}$, we refer to \eqref{eq:model} as the exact unlabeled sensing problem. 

\begin{figure}[h!]
    \centering
    \includegraphics[width = 0.61 \linewidth]{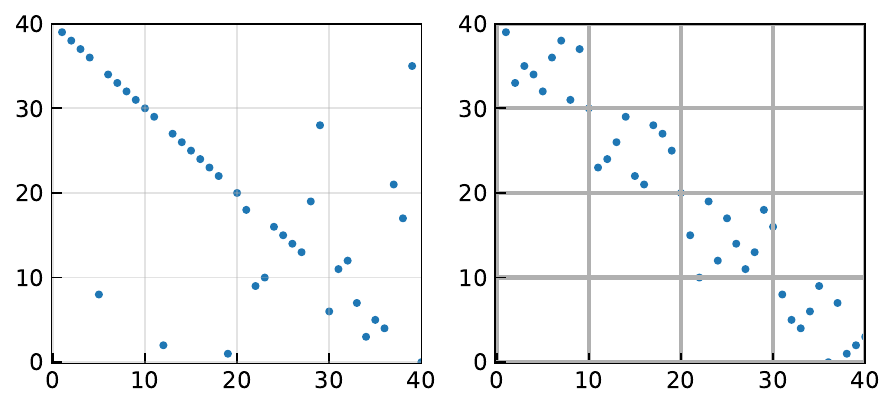}
\caption{\small{Left.  Sparse (or partially shuffled) permutation considered in \cite{snr,slawski_two_stage,zhang2019permutation,slawski2020sparse}, with number of shuffles $k=10$. Right. The $r$-local permutation structure considered in \cite{ojsp, icassp_r_local, wang2023regularization}, with block size $r=10$. In this paper, we propose a general algorithm for both permutation models. }}
 \label{fig:models}
\end{figure}

Since estimating $\X^*$ and $\P^*$ for a generic permutation is challenging, several works \cite{snr, slawski_two_stage, zhang2019permutation, slawski2020sparse, spherical, slawski2019linear} assume a $k$-sparse or partially shuffled permutation model, Figure \ref{fig:models}. A permutation matrix $\P^*_k$ is $k$-sparse if it has $k$ off-diagonal elements, i.e.,
$\langle \mathbf{I}, \P^*_k \rangle = n - k$, where $\langle \cdot, \cdot \rangle$ denotes the trace inner product. The $r$-local, or block diagonal, permutation model was first proposed in \cite{ojsp} and later considered in \cite{marano, icassp_r_local}. A permutation matrix $\P^*_r$ is $r$-local with $s$ blocks if $\P^*_r = \textrm{blkdiag}(\P_1, \dots, \P_s)$,
where $\textrm{blkdiag}(\cdot)$ denotes a block diagonal matrix. Fig. \ref{fig:models} illustrates the $k$-sparse and $r$-local permutation models. These two models are compared by information-theoretic inachievability in \cite{icassp_r_local}.

In \cite{icassp_r_local}, an alternating minimization algorithm was proposed for the unlabeled sensing problem. The algorithm  estimates  $\P^*$ and $\X^*$ by minimizing the following optimization program:
\begin{equation}
(\widehat \P,\widehat \X) =  \underset{\P \in \Pi_n, \X}{\argmin}\, F(\P,\X) = \lVert \Y-\P \B\X\rVert_{F}^{2}, \label{eq:obj} \end{equation}
where $\Pi_n$ denotes the set of $n\times n$ permutations and $\lVert \cdot \rVert_F^2$ denotes the squared Frobenius norm of a matrix, which is the sum of the squares of its entries. Given that \eqref{eq:obj} is a non-convex optimization problem, a crucial part of the alternating minimization algorithm in \cite{icassp_r_local} is the initialization. When the unknown permutation $\P^*_r$ is $r$-local, \cite{icassp_r_local} proposed the \textit{collapsed} initialization \eqref{eq:xhat_collapsed}. Let $\P_i$ denote each  $r_i \times r_i$ block of  $\P^*_r = \textrm{blkdiag}(\P_1,\cdots,\P_s)$. Let $\B_i \in \mathbb{R}^{r_i \times d}$ denote blocks of the measurement matrix $\B = [\B_1;\cdots;\B_s]$, where $;$ denotes vertical concatenation. Then, each block of measurements $\Y_i$ is expressed as $\Y_i = \P_i \B_i \X^*,  i \in [s]$. The shuffled measurements in each block  $[\P_i \B_i : \Y_i]$ are summed to extract $s$ unshuffled measurements $ [\mathbf{1}_{r_i}^{\top} \P_i \B_i : \mathbf{1}_{r_i}^{\top} \Y_i] \rightarrow [\tilde{\b}^\top_i : \tilde \y^\top_i]$, where $\mathbf{1}_{r_i} \in \mathbb{R}^{r_i}$ denotes the vector whose entries are all $1$. These $s$ measurements are then represented compactly in the \textit{collapsed} linear system of equations
$ \widetilde \B \X^* = \widetilde \Y \label{eq:collapsed_model}$, where $\widetilde \B \in \mathbb{R}^{s \times d}$ and $\widetilde \Y \in \mathbb{R}^{s \times m}$.  The  initialization for $r$-local $\P^*_r$ is the minimum-norm solution to the collapsed system and is given by:
\begin{equation}
    \widehat \X^{(0)}_r  = \widetilde\B^\dagger \widetilde\Y
    \label{eq:xhat_collapsed},
\end{equation}
where $\widehat \X^{(0)}_r = [\hat \x^{(0)}_1 \mid \dots \mid \hat \x^{(0)}_m]$. For the $k$-sparse problem, in this manuscript, we propose the following initialization:
\begin{equation}
\label{eq:ksparseInit}    
\widehat \X^{(0)}_k = \B^\dagger \Y.
\end{equation}
Note that the above initialization corresponds to the least square solution
of \eqref{eq:model} when $\P^*$ is the identity matrix. The initialization is motivated by the observation that, since $\P^*_k$ has $k$ off-diagonal elements, the identity matrix serves as a simple approximation of $\P^*_k$. The main goal of this manuscript is to characterize the effectiveness of the initializations $\X^{(0)}_r$ and $\X^{(0)}_k$ by upper bounding the initialization error for the exact unlabeled sensing problem.

\subsection{Contributions and outline}

This paper presents a theoretical analysis of the initialization for the exact unlabeled sensing problem under two permutation models. The key contributions of this manuscript are summarized as follows:

\begin{enumerate}
\item Assuming $\B$ is Gaussian, we show that the relative error $\displaystyle
\frac{\lVert \X^* - \widehat \X^{(0)}_r \rVert_F}{\lVert \X^* \rVert_F}
$ critically depends on the parameter $d - s$. In particular, Theorem \ref{thm:multiViewfixedXrandomB} establishes that the probability that the relative error exceeds $\sqrt{1 - s/d}$ decays at a sub-Gaussian rate. 

\item Assuming $\x^*$ is sub-Gaussian, we focus on the error $\displaystyle \left[ \sum_{j=1}^{m} \lVert \x^*_j - \hat \x^{(0)}_j \rVert_2 \right],$ which also depends critically on the parameter $d - s$. In particular, Theorem \ref{thm:multiViewRandomX} in this work shows that the probability that the error exceeds $O\big(\sqrt{d - s}\big)$ also decays at a sub-Gaussian rate.
\item For the $k$-sparse problem, we analyze the relative error $\displaystyle \frac{\lVert \X^* - \widehat \X^{(0)}_k \rVert_F^2}{\lVert \X^* \rVert_F^2}.$ Assuming $\B$ is Gaussian and sufficiently tall (see Definition \ref{defn:tall}), we show that the relative error depends on the parameter $k/n$.  Specifically, depending on the problem parameters, Theorem \ref{thm:kSparse} shows that the probability of the error exceeding $O(k/n)$ decays at a rate ranging from inverse linear to exponential.
\item We apply the alternating minimization algorithm in \cite{icassp_r_local} with the initializations \eqref{eq:xhat_collapsed} and \eqref{eq:ksparseInit}. Experiments results on real
datasets show superior performance compared to baseline methods.
\end{enumerate}

\paragraph{\textbf{Outline}:} The outline of the paper is as follows. Section 2 discusses related work. Section 3 presents the alternating minimization algorithms for the $r$-local and $k$-sparse permutation models. Section 4 covers the initialization analysis. Experimental results on real
datasets showing superior performance compared to baseline methods are given in Section 5. We conclude in Section 6.

\subsection{Notation}
Boldface lowercase letters denote column vectors and boldface uppercase letters denote matrices. The transpose of a vector $\mathbf{x}$ is noted by $\mathbf{x}^{\top}$. For a matrix $\mathbf{X}$, its transpose is denoted by $\mathbf{X}^{\top}$. $\mathbf{A}^\dagger$ denotes the Moore-penrose pseudoinverse of $\mathbf{A}$. 
$\Tr(\cdot)$ denotes the trace of a matrix. Given two matrices $\mathbf{A}$ and $\mathbf{B}$, $\langle \mathbf{A}\,, \mathbf{B} \rangle$ denotes the trace inner product, defined as $\Tr(\mathbf{A}^\top \mathbf{B})$. Given a vector $\mathbf{x}$, the $i$-th entry of $\mathbf{x}$ is denoted by $x_i$. $[\A ; \B]$ denotes the vertical concatenation of matrices $\A$, $\B$. 
$\lVert \cdot \rVert_p$ denotes the vector $p$-norm. $\lVert \cdot \rVert_F$ and $\lVert \cdot \rVert$ denote the Frobenius norm and the operator norm of a matrix, respectively. $\mathbf{1}_n \in \mathbb{R}^n$ denotes the vector of all ones. $\mathbf{I}$ denotes the identity matrix. $\real$ denotes the set of real numbers.  $\Pi_n = \{\Z \colon \Z \in \{0,1\}^{n \times n}, \Z\mathbf{1}_n = \mathbf{1}_n, \Z^\top \mathbf{1}_n = \mathbf{1}_n\}$ denotes the set of $n \times n$ permutation matrices. $\mathbb{E}(\cdot)$ denotes the expectation of a random variable in consideration. $C,C'$ denote absolute constants $ > 1$, and  $c,c'$ denote absolute constants $\leq 1$. $\argmin$ denotes the set of minima of an objective function in consideration. $\exp(\cdot)$ denotes the exponential function. 
We also summarize the frequently used notation in Table \ref{tbl:notation}.
\begin{table*}
\centering
\renewcommand{\arraystretch}{1.35} 
\begin{tabular}{|p{2.5cm}|p{8.25cm}|}
    \hline
    \textbf{Symbol} & \textbf{Description} \\
    \hline
    $\B \in \mathbb{R}^{n \times d}$ & Measurement matrix\\
    \hline
    $\X^* \in \mathbb{R}^{d \times m}$ &  Unknown matrix\\
    \hline
    $\x^*_j \in \mathbb{R}^{d}$ &  $j$-th column of $\X^*$, \, $\forall j \in [m]$\\
    \hline 
    $\P^* \in \{0,1\}^{n \times n}$ &  Unknown permutation matrix\\
    \hline
    $\P^*_k \in \mathbb{R}^{n \times n}$ &  Unknown $k$-sparse permutation matrix \\
    \hline
    $k$ & Number of shuffles $k$ s.t. $\langle \P^*_k,\mathbf{I} \rangle = n - k$ \\
    \hline
    $\P^*_r \in \{0,1\}^{n \times n}$ &  Unknown $r$-local permutation matrix\\
    \hline 
    $r_i \, \forall \, i \in [s]$ & Block sizes of $\P^*_r$ s.t. $\P^*_r = \mathrm{blkdiag}\, (\P^*_1, \cdots, \P^*_s)$\\
    \hline
    $\Pi_{r_i}$ & The set of $r_i \times r_i$ permutation matrices.\\
    \hline
    $s$ & Number of blocks in $r$-local $\P^*_r$\\
    \hline
    $[s]$ & Integers $\{1, \cdots, s\}$ \\
    \hline 
\end{tabular}
\caption{Table summarizing the frequently used notation used in this paper. See Figure 1 for definition and examples of $r$-local $\P^*_r$ and $k$-sparse  $\P^*_k$ permutations. The superscript $*$ in $\P^*,\P^*_r,\P^*_k,\X^*$ denotes that these are unknown problem parameters which are estimated by the proposed algorithm.}
\label{tbl:notation}
\end{table*}

\section{Related Work}

This section provides a concise overview of related work in unlabeled sensing theory and algorithms, inference problems involving unlabeled data, and applications of unlabeled sensing.

\subsection{Theory and algorithms}

We note that the problem in \eqref{eq:model} is referred to as the single-view (multi-view) unlabeled sensing problem for $m = 1$ ($m > 1$). For the single-view problem, we represent the problem as $\y = \P^* \B \x^* + \w$ where $\y$, $\x^*$, and $\w$ denote the  variables analogous to $\Y$, $\X^*$ and $\W$ respectively. The work in \cite{unnikrishnan2018unlabeled} formulated the single-view unlabeled sensing problem and established that $n = 2d$ measurements are both necessary and sufficient for the recovery of $\x^*$. Subsequent works \cite{Dokmanic, pananjady} generalized this result and developed an information-theoretic inachievability analysis. 

For single-view unlabeled sensing, algorithms based on branch and bound and expectation maximization are proposed in \cite{emiya, concave, header_free}, which are suitable for small problem sizes. A stochastic alternating minimization approach is introduced in \cite{abid2018stochastic}. For multi-view unlabeled sensing, the Levsort subspace matching algorithm was proposed in \cite{Levsort}, and a subspace clustering approach was presented in \cite{slawski2020sparse}. The works \cite{snr, slawski_two_stage, icml} propose methods based on bi-convex optimization, robust regression, and spectral initialization, respectively.

\subsection{Inference on unlabeled data}

The problem of unlabeled sensing has been studied in the context of estimation and detection from unlabeled samples. In \cite{wang2018signal}, the authors consider the problem of signal amplitude estimation and detection from unlabeled quantized binary samples. In the setup they consider, the ordering of the time indexes is unknown. The work in \cite{wang2018signal} proposes a maximum likelihood estimator to estimate the underlying signal amplitude and permutation, under certain structural assumptions on the quantization and signal profile. Furthermore, an alternating maximization algorithm is studied for the general estimation and detection problem.

In \cite{haghighatshoar2017signal}, the authors study signal recovery from unlabeled samples by considering a special case of unlabeled sensing, referred to as the unlabeled ordered sampling problem. In this problem, instead of estimating an unknown permutation matrix, the goal is to estimate an unknown selection matrix that preserves the order of the measurements. \cite{haghighatshoar2017signal} links this problem to compressive sensing and also proposes an alternating minimization algorithm for solving it. \cite{liu2018signal} also addresses the unlabeled ordered sampling problem, but with a focus on signal detection, rather than signal recovery as in \cite{haghighatshoar2017signal}. A dynamic programming algorithm is proposed therein to estimate the selection matrix. 

\cite{marano2019algorithms} considers the problem of signal detection, where the observations are independently drawn from a finite alphabet. Focusing on the inference problem, the work in \cite{marano2019algorithms} characterizes the information available in the unordered samples by studying a binary hypothesis test. Additionally, it provides a computationally efficient detector to address the detection problem. \cite{marano2020making} further studies the signal detection problem in the case of binary unlabeled observations. In the low-detectability regime, \cite{marano2020making} gives an analytical characterization of various statistical inference quantities, such as Chernoff information. In \cite{sutton2023identity}, the signal detection framework of unlabeled sensing is applied to decision-making in sensor networks, where sensors report measurements to a central node, but the measurements are imprecise or lack labels. \cite{sun2023quickest} considers a related problem, where sensors send their measurements to a central fusion center. Due to an anomaly in one of the sensors, the measurements at the fusion center are assumed to be unordered, and \cite{sun2023quickest} explores the problem of detecting the anomaly with minimal detection delay.

\subsection{Applications of unlabeled sensing}
\label{subsec:linkedLinearRegression} 
\begin{figure}[h!]
    \centering
    \begin{subfigure}{0.5\linewidth}
    \includegraphics[height = 3.0cm]{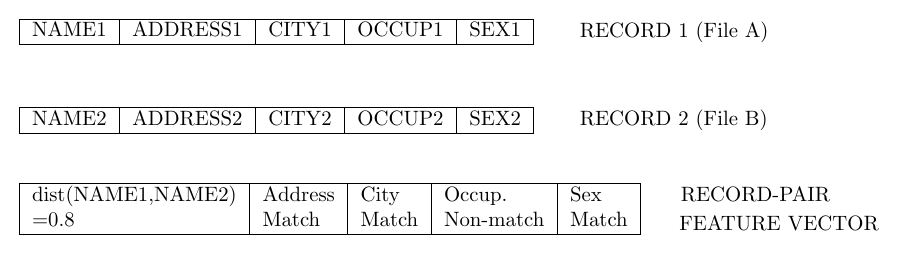}
    \end{subfigure}
   \begin{subfigure}{0.45\linewidth}        
   \centering
    \includegraphics[width=0.5\linewidth]{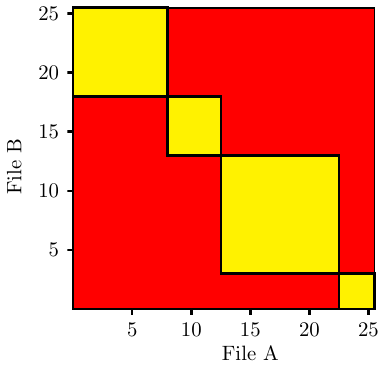}      
        \end{subfigure}
\caption{\textit{Record linkage}  with \textit{Blocking} \cite{murray2016probabilistic},  \cite{graph_rl}  assigns records in File A to records in File B, upto blocks. For example, records matching on identifiers `city', `occupation' etc (left)  are assigned to  the same block (right). Linked linear regression  \cite{lahiri2005regression} fits a regression model on such block-permuted data. See Section \ref{sec:results} for results on real datasets. The figure on the left is adapted from Figure 1 in \cite{graph_rl} and the figure on the right is adapted from Figure 1 in \cite{murray2016probabilistic}.  }
\label{fig:exmplrLcl}
\end{figure}

The linked linear regression problem \cite{murray2016probabilistic, lahiri2005regression,han2019statistical}, also called regression analysis with linked data, is to fit a linear model by minimizing over $\x$ and $\P$ the objective $\lVert \y - \P\B\x\rVert_2$, where $\P$ is an  $r$-local permutation and $\x$ is the regression vector. For example, consider the  statistical model where the weight depends linearly on age and height. Let $\y \in \mathbb{R}^n$ contain the weights of $n=10$ individuals, 5 males, 5 females. Let $\B \in \mathbb{R}^{n \times d}$, with $d=2$ contain the age and height values. Each record (weight, age, height) is known to belong to a male or a female individual. Letting the first (second) block of 5 rows of $\y,\B$ contain the records of male (female) individuals, the unknown $r$-local permutation, $r=5$, assigns each weight value (row of $\y$) to its corresponding age, height (row of $\B)$. Detailed references describing record linkage with blocking are \cite{kim2012regression}, Section 1.1 and \cite{wang2022regression}, Section 2; also, see 
 Fig. \ref{fig:exmplrLcl}. Experimental results on real datasets  are given in Section \ref{sec:results}. Other applications of unlabeled sensing include the pose and correspondence problem \cite{pananjady} and  $1$-d unassigned distance geometry  \cite{icassp_r_local}. 
In \cite{wang2020target}, the authors 
apply the unlabeled sensing framework to the sensor network localization problem.  We also note a recent work that extends the unlabeled sensing theory to multi-channel signals, leading to highly structured unlabeled sensing problems \cite{koka2024shuffled}. In \cite{koka2024shuffled}, beyond theoretical analysis of the structured problem, the model is applied to a real-world application in calcium imaging.

 \subsection{Technical background}

This section provides definition of sub-Gaussian and sub-exponential random vectors, and also summarizes the key concentration inequalities used throughout this paper. Let $\mathbf{Z}^* \in \mathbb{R}^{d \times m}$ be a matrix such that the columns $\mathbf{z}^*_j$ are  independent, zero-mean sub-Gaussian random vectors, with sub-Gaussian norm $K$. It follows then that 
\begin{equation}
    \mathbb{E} [\exp (\alpha^\top \x_j^*) ] \leq \exp(\lVert \alpha \rVert_2^2 K^2/2 ) \, \forall \,\,\, \alpha \in \mathbb{R}^d \text{ and } \forall j\in \{1,...,m\}.
    \label{eq:defsubG}
\end{equation}
A random variable $X$ is sub-exponential  with sub-exponential norm $K_{d-s}$ if
\begin{equation}
\Pr[\lvert X \rvert \geq t] \leq C\exp(-t/K_{d-s}),
\label{eq:defsubExp}
\end{equation}
where $C \geq 1, t \geq 0$ and $K_{d-s} \geq 0$. 
\begin{theorem}[Hanswon Wright Inequality, Theorem 2.1 in \cite{hsu2012tail}] Let $\mathbf{\Sigma} = \mathbf{A}^{\intercal}\mathbf{A}$ be a positive semi-definite matrix. Let $\x = (x_1,\cdots,x_d)$ be a zero-mean sub-Gaussian random vector, i.e., for $\alpha \in \mathbb{R}^d$, $K \geq 0$
\begin{equation*}
    \mathbb{E} [\exp (\alpha^\intercal \x^*) ] \leq \exp(\lVert \alpha \rVert_2^2 K^2/2 ).
\end{equation*}
For $t \geq 0$, 
\begin{equation}
\Pr[\lVert \mathbf{A}\x \rVert_2^2 \geq  K^2  ( \Tr(\mathbf{\Sigma}) + 2 \sqrt{\Tr(\mathbf{\Sigma}^2)t} + 2 t\lVert \mathbf{\Sigma} \rVert) ] \leq e^{-t}. \label{eq:PSD_HWI}
\end{equation}
\end{theorem}
\begin{lemma}[Johnson-Lindenstrauss Lemma, Lemma 5.3.2 in \cite{vershynin2018high}]
Let $P$ be a projection in $\mathbb{R}^p$ onto a uniformly distributed random $q$-dimensional subspace. Let $z \in \mathbb{R}^p$ be a fixed point and $t > 0$. Then, with probability at least  $1 - 2\exp(-c t^2 q)$, 
\begin{equation}(1 - t) \sqrt{\frac{q}{p}} \lVert z \rVert_2 \leq \lVert P z \rVert_2 \leq (1+t) \sqrt{\frac{q}{p}}\lVert z \rVert_2. \label{eq:JL} \end{equation}
\end{lemma}
\begin{theorem}[Hoeffding's inequality, Theorem 2.6.2 in \cite{vershynin2018high}] Let $X_1,\cdots,X_m$ be independent, sub-Gaussian random variables. Then, for every $t \geq 0$, 
\begin{equation}
    \Pr[\lvert \sum_{i=1}^{i=m} X_i  - \mathbb{E}[X_i] \rvert \geq t] \leq 2 \exp \Big(\frac{-c't^2}{\sum_{i=1}^{i=m} \lVert X_i \rVert_{\varphi_2}^2}\Big),
    \label{eq:Hoeffding}
\end{equation}
where $\lVert X_i \rVert_{\varphi_2}$ denotes the sub-Gaussian norm of $X_i$ and $c'$ is an absolute constant.
\end{theorem}
\begin{lemma}[Tail inequality for $\chi^2_D$ distributed random variables, Lemma 1 in \cite{10.1214/aos/Laurent}] Let $Z_D$ be a $\chi^2$ statistic with $D$ degrees of freedom. For any positive $t$,
\begin{align}
    \Pr [Z_D \geq D + 2\sqrt{Dt} + 2 t] & \leq e^{-t}, \label{eq:chiSquareUpper}\\
    \Pr [Z_D \leq D -  2\sqrt{Dt}] & \leq e^{-t}. \label{eq:chiSquareLower}
\end{align}
\end{lemma}

\section{Algorithm}
In this section, we briefly summarize the alternating minimization algorithm first proposed in \cite{icassp_r_local}. To estimate the  $\P^*$ and $\X^*$ in the optimization objective \eqref{eq:model}, the alternating minimization (AltMin)  updates for \eqref{eq:obj}  are
\begin{align}
    &\P^{(t)}  =  \argmin_{\P \in \Pi_n} \langle -\Y {\X^{(t)}}^{\top}\B^{\top},\P \rangle, \label{eq:P_update}\\
    &\X^{(t+1)} = {(\P^{(t)}\B)}^{\dagger} \Y,
    \label{eq:x_update}
\end{align}
where $\P^{(t)}$ and $\X^{(t)}$ denote the estimate of $\P^*$ and $\X^*$ at the t-th iteration respectively. The initialization to  \eqref{eq:P_update}, \eqref{eq:x_update}, a linear assignment problem and a least squares problem,  has to be chosen carefully because \eqref{eq:obj} is a non-convex optimization problem. After initialization, we alternate \eqref{eq:P_update}, \eqref{eq:x_update} and use the relative change in the objective value $F(\X,\P)$ as the stopping criterion. For $r$-local $\P^*_r$, the $\P$-update \eqref{eq:P_update} decouples along the blocks of $\P^*_r$, see Algorithm \ref{alg:AltMin_rLcl}.  For $k$-sparse $\P^*_k$, see Algorithm \ref{alg:AltMin_ksparse}. 

\begin{algorithm}[ht!]
\caption{AltMin for $r$-local $\P^*_r$}
\label{alg:AltMin_rLcl}
\begin{algorithmic}[1]
\REQUIRE Convergence threshold $\epsilon$, $\Y$, $\B$ and block sizes $r_1,r_2,..,r_s$ of $\P^*_1, \P^*_2,...,\P^*_s$ respectively. 
        \STATE{$\widehat \X \gets$ collapsed initialization in \eqref{eq:xhat_collapsed} }
        \STATE{$\widehat \Y \gets \B\widehat\X$}
        \WHILE {$\textrm{relative change} >  \epsilon $} 
            \FOR {$i \in 1 \cdots s$} \,\,\, //$s$ \text{is the number of blocks}        
                \STATE{$\widehat{\P}_i \gets {\argmin}_{\P_i \in \Pi_{r_i}} - \langle  \Y_i \widehat{\Y}_i^{\top},  \P_i   \rangle$}
            \ENDFOR
        \STATE {$\widehat{\mathbf{\P}} \gets  \textrm{blkdiag} (\widehat{\P}_{1},\cdots,\widehat{\P}_{s})$} 
        \State{$\widehat{\X} \gets \B^{\dagger}  \widehat{\P}^{\top} \Y$} 
        \State{$\widehat \Y \gets \B \widehat \X$}
        \ENDWHILE
        \STATE {\bfseries Return} $\widehat \P, \widehat \X$
\end{algorithmic}
\end{algorithm}
\begin{algorithm}[ht!]
\caption{AltMin for $k$-sparse $\P^*_k$}
\label{alg:AltMin_ksparse}
\begin{algorithmic}[1]
\REQUIRE convergence threshold $\epsilon$, $\Y$, $\B$
        \STATE{$\widehat \Y \gets \Y$}
        \WHILE {$\textrm{relative change} >  \epsilon $} 
        \STATE{$\widehat \P \gets {\argmin}_{\P \in \Pi_n} - \langle  \Y \widehat{\Y}^{\top},  \P   \rangle$}
        \State{$\widehat{\X} \gets \B^{\dagger}  \widehat{\P}^{\top} \Y$} 
        \State{$\widehat \Y \gets \B \widehat \X$}
    \ENDWHILE
        \STATE {\bfseries Return} $\widehat \P, \widehat \X$
\end{algorithmic}
\end{algorithm}

\section{Initialization analysis}

In this section, we derive upper bounds for the initialization error of the alternating minimization algorithm applied to solve the exact unlabeled sensing problem (\eqref{eq:model} with $\mathbf{W} = \mathbf{0}$), using the initializations in \eqref{eq:xhat_collapsed} and \eqref{eq:ksparseInit}.

\subsection{Analysis for $r$-local permutation}
 \label{subsec:initialization_rLocal}
We consider the initialization in \eqref{eq:xhat_collapsed}. Let $\widetilde \B = \widetilde\U \S \widetilde \V^\top$ denote the compact singular value decomposition of $\widetilde \B$. Using this in \eqref{eq:xhat_collapsed}, we obtain $\widetilde\B^\dagger = \widetilde\V\S^{-1}\widetilde\U^\top$, where $\widetilde \B = \widetilde\U \S \widetilde \V^\top$. Using this substitution, it can be shown that the initialization error $\lVert \X^* - \widehat \X^{(0)}_r \rVert_F$  is the projection of $\X^*$ onto the the orthogonal complement of the row space of $\widetilde \B$, as follows:
\begin{equation}
    \lVert \X^* - \widehat \X^{(0)}_r \rVert_F = \lVert \X^* - \widetilde\B^\dagger \widetilde\B \X^*    \rVert_F =   \lVert ( \mathbf{I} - \widetilde \V \widetilde \V^\top   )\X^* \rVert_F.
    \label{eq:row_proj}
\end{equation}
Recall that the size of $\widetilde\B$ is $s\times d$. If $s\ge d$, and assuming that $\widetilde\B$ has rank $d$, it can be verified that $\widehat\X^{(0)}_r = \X^*$. 
For instance, for $s\ge d$, if the entries of $\B$ are drawn independently from a continuous distribution, the rank of $\widetilde \B$ is $d$ with high probability. Given that, in this section, we upper bound the error in initialization
for the under-determined $s < d$ case. Similar analysis for under-determined systems have been studied in the sketching literature \cite{woodruff2014sketching}, but these results are not applicable here as our sub-sampling strategy via the collapsing initialization is deterministic.

The first key theoretical result is Theorem \ref{thm:multiViewfixedXrandomB}. This theorem considers a Gaussian matrix $\B \in \mathbb{R}^{n \times d}$, with $\X^* \in \mathbb{R}^{d \times m}$ assumed to be a fixed but unknown matrix. To upper bound the relative error $\displaystyle \frac{\lVert  \X^* - \widehat \X^{(0)}_r \rVert_F}{\lVert \X^* \rVert_F}$, we apply the Johnson–Lindenstrauss lemma. The key insight is that the relative error depends on the parameter $d - s$. Specifically, the probability that the relative error exceeds $\sqrt{1-\frac{s}{d}}$ decays at a sub-Gaussian rate. This implies that the initialization in \eqref{eq:xhat_collapsed} improves as the number of blocks $s$ in the $r$-local model increases.

\begin{theorem} \label{thm:multiViewfixedXrandomB}
Let $\X^* \in \mathbb{R}^{d \times m}$ be the fixed unknown  matrix and let $\widehat \X^{(0)}_r$ be as defined in \eqref{eq:xhat_collapsed}. Consider the exact unlabeled sensing problem. Assuming Gaussian $\B \in \mathbb{R}^{n \times d} $ and block-diagonal $\P^*_r$ with $s$ blocks,  for $\log m \leq ct^2(d-s)$, $s < d$, and $t \geq 0$, 

\begin{equation}
\Pr \bigg [\frac{\lVert  \X^* - \widehat \X^{(0)}_r \rVert_F}{\lVert \X^* \rVert_F} \geq  (1 + t)\sqrt{\frac{d-s}{d}} \bigg] \leq 2 \exp (-c(d-s) t^2).
\label{eq:randomBfixedXMultiView}
\end{equation}
\end{theorem}
\begin{proof} 
The error \eqref{eq:row_proj} is the column-wise projection of $\X^* \in \mathbb{R}^{d \times m}$ onto a $(d - s)$-dimensional uniformly random subspace \eqref{eq:row_proj}, which can be bounded by the JL Lemma \eqref{eq:JL}. 
\end{proof}
In Theorem \ref{thm:multiViewfixedXrandomB}, we considered a Gaussian matrix $\B$ and a fixed, unknown matrix $\X^*$. In the following, we fix the measurement matrix $\B$ and introduce randomness in $\X$. Before discussing our main result for this setting, we provide motivation for considering this scenario. One motivating application is the unassigned distance geometry problem (uDGP) \cite{duxbury2016unassigned, duxbury2022unassigned, Lemke2003}, which involves recovering the configuration of points from a list of distances. We focus on the 1-dimensional uDGP, which can be framed as a structured unlabeled sensing problem, as described in \cite{icassp_r_local}, where $\mathbf{x}^*$ represents the sought-after positions of the points on the 1D line. In this case, the matrix $\B$ is deterministic (see equation (3) of \cite{icassp_r_local}). For this setting, without any assumptions on the underlying points $\mathbf{x^*}$, and assuming that the underlying permutation is $r$-local, the collapsed initialization may be suboptimal. To illustrate this, recall the error in the collapsed estimate: 
$\lVert ( \mathbf{I} - \widetilde \V \widetilde \V^\top ) \mathbf{x^*} \rVert_F$. For any $\mathbf{x^*}$ orthogonal to the subspace spanned by $\widetilde \V$, the collapsed estimate will be the zero vector. 
This shows that the following upper bound on the initialization error 
$\lVert \hat \x_r^{(0)} - \mathbf{x^*} \rVert \leq \lVert \mathbf{I} - \widetilde \V \widetilde \V^\top \rVert \lVert \mathbf{x^*} \rVert= \lVert \mathbf{x^*} \rVert$, holds with equality for $\mathbf{x^*}$ orthogonal to the subspace spanned by the columns of $\widetilde \V$. The implication of the above discussion is that some structural assumption on the underlying $\mathbf{x^*}$ is necessary to obtain suitable bounds on the initialization error.

We next consider the squared error in the initialization $\lVert \A \x^* \rVert^2$ where $\A = \mathbf{I} - \widetilde \V \widetilde \V^\top$. Note that $\A$ is an orthogonal projection matrix onto the orthogonal complement of the subspace spanned by $\widetilde \V$. Given that $\widetilde \V \in \mathbb{R}^{d \times s}$ and we are considering the underdetermined regime $s < d$, the rank of $\A$ is $d - s$. Therefore, the analysis of the error is equivalent to analyzing the quadratic form $\lVert \A \x^* \rVert^2$. One common approach to control the norm of a random vector is to assume that $\x^*$ is a random vector with independent, sub-Gaussian coordinates (see, for example, Theorem 3.1.1 in \cite{vershynin2018high}). However, we note that this assumption does not hold after applying the projection operator $\A$. A more suitable tool in this case is the Hanson-Wright inequality. In fact, the quadratic form in our case is special due to the fact that $\A$ is a projection operator, allowing us to leverage its linear algebraic properties. To apply the Hanson-Wright inequality, it suffices to assume that $\x^*$ is sub-Gaussian with mean zero, without requiring the independence of its coordinates. With this assumption, we study the error 
$\displaystyle \left[ \sum_{j=1}^{m} \lVert \x^*_j - \hat \x^{(0)}_j \rVert_2 \right]$. Theorem \ref{thm:multiViewfixedXrandomB} shows that this error depends critically on the parameter $d - s$. In particular, Theorem \ref{thm:multiViewRandomX}  demonstrates that the probability of the error exceeding $O\left( \sqrt{d - s} \right)$ decays at a sub-Gaussian rate.

\begin{theorem}
\label{thm:multiViewRandomX}
Let $\X^* \in \mathbb{R}^{d \times m}$ be the unknown matrix such that the columns $\x^*_j$ are  independent, zero-mean sub-Gaussian random vectors, with sub-Gaussian norm $K$. Let $\widehat \X^{(0)}_r = [\hat \x^{(0)}_1 \mid \cdots \mid \hat \x^{(0)}_m]$ be as defined in \eqref{eq:xhat_collapsed}. Consider the exact unlabeled sensing problem. For any fixed measurement matrix $\B$ and block-diagonal $\P^*_r$ with $s$ blocks, $s < d$  and $t \geq 0$, 
 \begin{equation}
      \hspace{-0.5em}\Pr \big[\sum_{j=1}^{j=m}\lVert \x^*_j - \hat \x_{j}^{(0)} \rVert_2  - m C_{d-s} \geq t \big]  \leq 2 \exp \bigg( \frac{-ct^2}{mK_{d-s}}\bigg) , \label{eq:Thm2Main}
\end{equation}
where $C_{d-s} =  K(d-s + \frac{5}{2}\sqrt{d-s} + 2)^{\frac{1}{2}}$, $K_{d-s} = K^2\sqrt{d - s}$.
\end{theorem} 

The strategy for the proof of Theorem \ref{thm:multiViewRandomX}  is to first use the Hanson-Wright inequality to control the quadratic form $\lVert \A \x^* \rVert^2$. Once that is established, we use Lemma \ref{lem:HWI} to define a modified random variable that is sub-exponential. The rest of the technical proof uses the Hoeffding bound on the modified random variable and standard techniques to relate the concentration inequality of $\lVert \mathbf{z} \rVert_2^2$ to $\lVert \mathbf{z} \rVert_2$ \cite{vershynin2018high}.
For the complete proof, see the Appendix \ref{sec:AppndxA}.

\begin{lemma}\label{lem:HWI} 
Let $\hat \x^{(0)}_r$ be as in \eqref{eq:xhat_collapsed},  $\x^* \in \mathbb{R}^d$ be a zero-mean sub-Gaussian random vector,  $\B$ be a fixed measurement matrix, and $\P^*_r$ be a fixed block-diagonal permutation with $s$ blocks. Let $\x_{\textrm{err}}$ denote the random variable $\x_{\textrm{err}} \equiv \lV  \x^* - \hat \x^{(0)}_r \rV_2^2$. 
\begin{enumerate}
    \item  For $s < d$ and $t \geq 0$,
    \begin{equation}\Pr [\x_{\text{err}}  -  C_{d-s} \geq t ] \leq  \exp({-ct}/{K_{d-s}}), \label{eq:random_x}\end{equation}
    where $K_{d-s} = K^2\sqrt{d-s}$,  $C_{d-s} = K^2(d-s +\frac{1}{2}\sqrt{d-s}).$  
    \item  Define the random variable $\tilde \x_{\text{err}}$ as follows:  
\begin{empheq}[left={\tilde \x_{\text{err}}=\empheqlbrace}]{align}  
C_{d-s} & \quad \x_{\text{err}} \leq C_{d-s} \label{eq:lower_tail} \\
\x_{\text{err}} & \quad \x_{\text{err}}   > C_{d-s}.     \label{eq:upper_tail}    
        \end{empheq} 
The random variable $\tilde \x_{\text{err}}$ is a sub-exponential random variable with sub-exponential norm $\lVert \tilde \x_{\text{err}} - C_{d-s} \rVert_{\varphi_1} = CK_{d-s}$. In addition, $\tilde \x_{\text{err}} \geq \x_{\text{err}}$, and  $\forall \, t > 0$,
    \begin{equation}
    \Pr[\tilde \x_{\text{err}}  - C_{d-s} \geq t] =  \Pr [\x_{\text{err}}   - C_{d-s} \geq  t].
    \end{equation}
\end{enumerate}
\label{lem:subExp}
\end{lemma}
\begin{proof}
We prove each part separately.
\begin{enumerate}
    \item 
 To derive \eqref{eq:random_x},  set $\A = (\mathbf{I} - \widetilde \V \widetilde \V^\top)$ in \eqref{eq:PSD_HWI},  where $\widetilde \V \in \mathbb{R}^{s \times d}$ denotes the basis for the row space of the collapsed matrix $\widetilde \B \in \mathbb{R}^{s \times d}$ in \eqref{eq:xhat_collapsed}. It follows that $\x_{\text{err}} \equiv \lV  \x^* - \hat \x^{(0)}_r \rV_2^2 = \lV \A\x^* \rV_2^2$.  Here, $\A \in \mathbb{R}^{d \times d}$ is  a $(d-s)$-dimensional projection matrix, and $\mathbf{\Sigma} = \A^\top\A = \A \A = \A$. Moreover,  $\Tr (\mathbf{\Sigma}) = \Tr (\mathbf{\Sigma}^2) = \Tr(\A) = d - s$ and the operator norm satisfies $\lVert \mathbf{\Sigma} \rVert = 1$. We now apply Hanson-Wright inequality in \eqref{eq:PSD_HWI}, we obtain:
\begin{equation}
\label{eq:hanson_1}
\Pr[\lVert \mathbf{A}\x \rVert_2^2 \geq  K^2  (d-s + 2 \sqrt{(d-s)t} + 2t]  \leq e^{-t}.
\end{equation}
The next step will be to upper bound the term $2\sqrt{(d-s)t}$. To do that, we rely on following
inequality: $2\sqrt{bt}\le 2t+\frac{1}{2}b$, for $b\ge 1$ and $t\ge 0$. To verify this, note that this inequality is equivalent to checking whether $4t\le 4t^2a+2ta+a/4$ holds. 
\begin{itemize}
    \item For $a=1$, this reduces to $0\le (2t-1/2)^2$, which holds for all $t\ge 0$. 
    \item For $a\ge 2$, we check $4t\le 4t^2a+2ta+1/4$ which is true for $t\ge 0$. 
    \end{itemize}
Using this inequality with $b=d-s$, we obtain $2 \sqrt{(d-s)t} \le 2t(d-s)+\frac{1}{2}(d-s)$. Substituting this bound in \eqref{eq:hanson_1}, we simplify the inequality as follows:
\begin{equation}
\label{eq:hanson_2}
\Pr[\lVert \mathbf{A}\x \rVert_2^2 \geq  K^2  (d-s + 2t(d-s)+\frac{1}{2}(d-s) + 2t]  \leq e^{-t}.
\end{equation}
The above inequality is equivalent to:
\begin{equation}
\Pr[\lVert \mathbf{A}\x \rVert_2^2 -K^2  (d-s + \frac{1}{2}(d-s)\ge 2K^2t((d-s)+1)]  \leq e^{-t}.
\end{equation}
A change of variable, $u=2K^2t((d-s)+1)$, and minor algebraic manipulation yields the final desired result.

\textbf{Remark:} To derive Theorem \ref{thm:multiViewRandomX}, we first show that  $\x_{\text{err}}$ is a sub-exponential random variable. It cannot be concluded from \eqref{eq:random_x} that the shifted random variable $\x_{\text{err}} - C_{d-s}$ is sub-exponential because the lower tail probability $\Pr[\x_{\text{err}} - C_{d-s} \leq 0]$
is not bounded. Since our goal is to upper bound the probability that the error \textit{exceeds} a certain value,  we define the non-negative sub-exponential random variable $\tilde\x_{\text{err}}- C_{d-s}$, which upper bounds the lower tail as $\{\x_{\text{err}} \leq C_{d-s} \} = C_{d-s} $,  see  \eqref{eq:lower_tail}. We will use the definition of $\tilde \x_{\text{err}}$ given in  \eqref{eq:lower_tail},  \eqref{eq:upper_tail}. 

\item Conditioning on the two complementary events in \eqref{eq:lower_tail}, \eqref{eq:upper_tail}, by the law of total probability, $\forall \ t > 0$, we have
\begin{flalign}
    &\Pr[\tilde \x_{\text{err}} - C_{d-s}  \geq t] \nonumber\\
    &=\Pr[\tilde \x_{\text{err}} - C_{d-s} \geq t \mid \x_{\text{err}}   \leq C_{d-s}]  \Pr[\x_{\text{err}}   \leq C_{d-s}]+ \Pr[\tilde \x_{\text{err}} - C_{d-s} \geq t \mid \x_{\text{err}} > C_{d-s}]  \Pr[ \x_{\text{err}} > C_{d-s}] \nonumber \\
    & =  \Pr[\tilde \x_{\text{err}}- C_{d-s} \geq t \mid \x_{\text{err}} > C_{d-s}]  \Pr[\x_{\text{err}} > C_{d-s}]   \label{eq:interm_step} \\
    & = \Pr[\x_{\text{err}}- C_{d-s} \geq t \mid \x_{\text{err}} > C_{d-s}]  \Pr[\x_{\text{err}} > C_{d-s}]\label{eq:condE2}\\ 
    & = \Pr[\x_{\text{err}} - C_{d-s} \geq t] \label{eq:claimProof},
\end{flalign}
 \eqref{eq:interm_step} follows from \eqref{eq:lower_tail}; specifically, $\Pr[\tilde \x_{\text{err}} - C_{d-s} \geq t \mid \x_{\text{err}} - C_{d-s}] = 0 \ \forall \ t > 0$. \eqref{eq:condE2} follows from \eqref{eq:upper_tail}. \eqref{eq:claimProof} follows from 
noting that the event $\{\x^* \mid \x_{\text{err}}  \geq  C_{d-s} + t \}$ is a subset of $\{\x^* \mid \x_{\text{err}}  >  C_{d-s} \}$. 
From \eqref{eq:claimProof}, for $t > 0$
\begin{align}
& \Pr[\x_{\text{err}} - C_{d-s}  \geq t ] =   \Pr [  \lvert \tilde \x_{\text{err}}  - C_{d-s} \rvert \geq t ].
\label{eq:step2}
\end{align}
$\eqref{eq:step2}$ follows from noting that $\tilde \x_{\text{err}} - C_{d-s} = \lvert \tilde \x_{\text{err}}  - C_{d-s}  \rvert$. Substituting \eqref{eq:random_x} in \eqref{eq:step2}, for $t > 0$,
\begin{equation} 
    \Pr [  \lvert \tilde \x_{\text{err}}  - C_{d-s} \rvert   \geq t ] \leq\exp\Big(\frac{-ct}{K_{d-s}}\Big).    \label{eq:extraStep}
\end{equation}
Since $\exp(0) = 1$, \eqref{eq:extraStep} holds for $t \geq 0$. In \eqref{eq:extraStep}, we have verified the definition in \eqref{eq:defsubExp} for $\tilde \x_{\text{err}} - C_{d-s}$ with $K_{d-s} = 2K^2(\sqrt{d-s}+1) = CK^2\sqrt{d-s}$. By proposition  2.7.1 (a), (d), definition 2.7.5 in  \cite{vershynin2018high} and \eqref{eq:extraStep}, the sub-exponential norm is $\lVert \tilde \x_{\text{err}} - C_{d-s} \rVert_{\varphi_1} = CK_{d-s}$.
\end{enumerate}
\end{proof}
\subsection{Analysis for $k$-sparse permutation}
\label{subsec:ksparse_init}

In this section, we analyze the initialization step when the underlying permutation is $k$-sparse. Specifically, we study the relative error 
$\frac{\lVert \X^* - \widehat \X^{(0)}_k \rVert_F^2}{\lVert \X^* \rVert_F^2}$. Assuming that $\B$ is a ``tall" Gaussian matrix, Theorem \ref{thm:kSparse} establishes that this error depends on the parameter $k/n$, which represents the proportion of shuffled entries relative to the total. More precisely, Theorem \ref{thm:kSparse} shows that the probability of the error exceeding $O(k/n)$ decays at a rate ranging from inverse linear to exponential.
Central to Theorem \ref{thm:kSparse} is the assumption that the measurement matrix $\B$ is a ``tall" Gaussian matrix, which we define precisely below.
\begin{defn}[\cite{rudelson2009smallest}]
   A Gaussian matrix $\B \in \mathbb{R}^{n \times d}$ is considered ``tall" if the aspect ratio $\lambda = d/n$ satisfies $\lambda < \lambda_0$ for some sufficiently small constant $\lambda_0 > 0$. In that case, we have $\Pr[\sigma_{\min}(\B) \geq c \sqrt{n}] \leq e^{-c n}$, where $c$ is an absolute constant that depends on $\B$. 
   \label{defn:tall}
\end{defn}

\begin{theorem} Let $\P^*_k$ be the fixed unknown $k$-sparse permutation matrix with $\langle \mathbf{I},\P^*_k \rangle = n-k$. Let $\X^*\in \mathbb{R}^{d\times m}$ be the fixed unknown matrix. Let $\B$ be a ``tall" Gaussian measurement matrix $\B$ as defined in Definition \ref{defn:tall}. For $\frac{2C\log(m)}{\sqrt{k}}\le \epsilon\le \frac{Ccn}{\sqrt{k}}$, where $C=4(1 + \sqrt{3})$, and $1\le k\le n$, we have
\begin{equation}
    \Pr\bigg[ \frac{\lVert \X^* - \widehat \X^{(1)} \rVert_F^2}{\lVert \X^* \rVert_F^2}  \geq  \left(\frac{2}{c^2}+\epsilon\right)\frac{k}{n}  \bigg] \leq 8\exp\left(-
    \frac{\epsilon \sqrt{k}}{2C}\right).
    \label{eq:relErrkSprse}
\end{equation}
\label{thm:kSparse}
\end{theorem}
For the proof of \eqref{eq:relErrkSprse}, see Section \ref{sec:AppndxA} in the Appendix. \\
\textbf{Remark:} In the above theorem, we consider the error bound at the lower bounds and upper bounds of $\epsilon$. When $\epsilon=\frac{2C\log(m)}{\sqrt{k}}$, it can be shown that the failure probability is $\frac{8}{m}$. When $\epsilon = \frac{Ccn}{\sqrt{k}}$, the failure probability is $\exp(-cn)$. \eqref{eq:relErrkSprse} gives a  ($<1$) relative error bound when $k$ grows slowly with $n$, for example $k = n^\beta$, $\beta < 1$.\\

\section{Results}
\begin{figure*}\renewcommand{\arraystretch}{1.3}
    \centering
    \begin{subfigure}[t]{0.333\textwidth}
    \begin{center}
    {\includegraphics[width = 1.0 \linewidth]{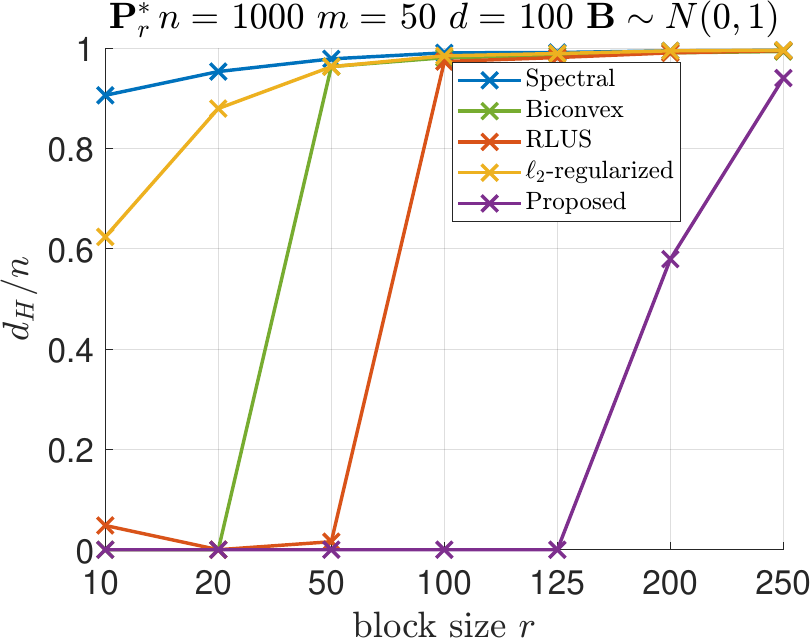}}
    \caption{}
    \label{subfig:rLocalSynth}
    \end{center}
    \end{subfigure}
    \begin{subfigure}[t]{0.32\textwidth}
    \begin{center}
    {\includegraphics[width = 1.00 \linewidth]{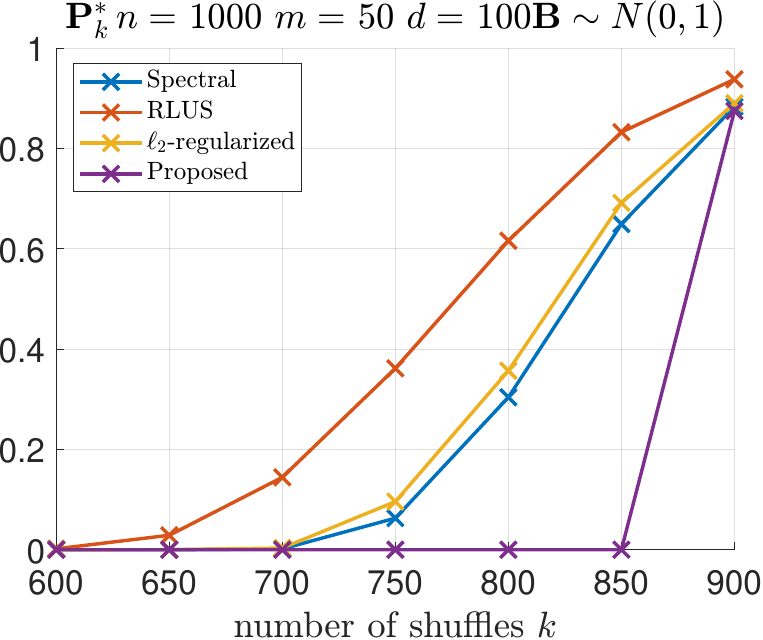}}
    \caption{}
    \label{subfig:randnB}
    \end{center}
    \end{subfigure}
    \begin{subfigure}[t]{0.32\textwidth}
    \begin{center}
    {\includegraphics[width = 1.00 \linewidth]{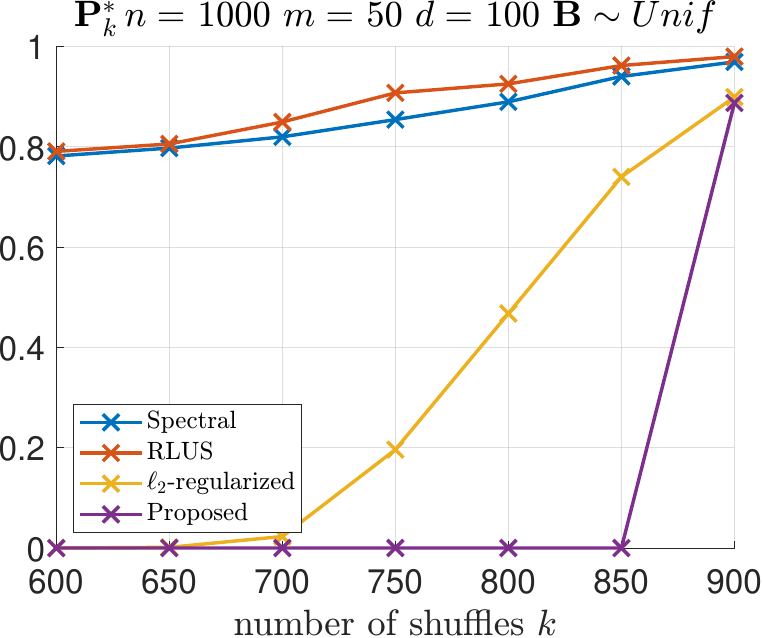}}
    \caption{}
    \label{subfig:randB}
    \end{center}
    \end{subfigure}
    \begin{subfigure}[t]{0.32\textwidth}
    \begin{center}
    {\includegraphics[width = 1.00 \linewidth]{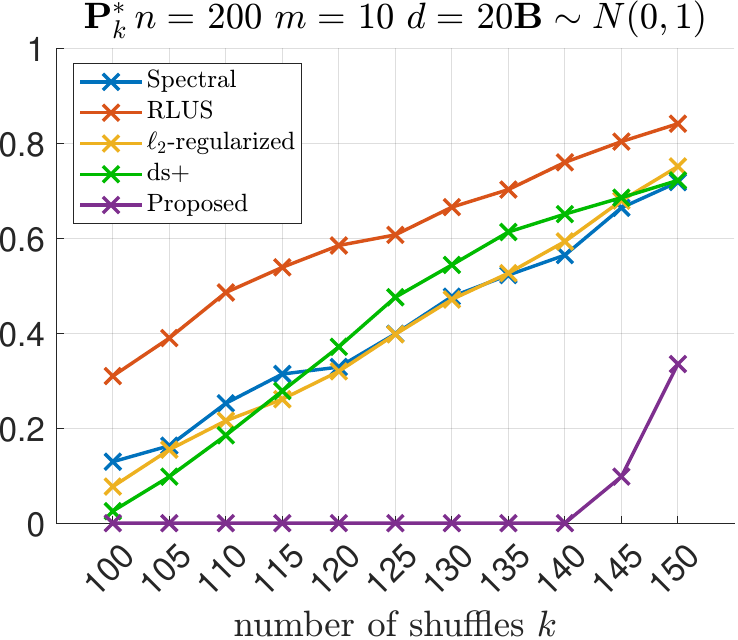}}
    \caption{}
    \label{subfig:ds+}
    \end{center}
    \end{subfigure}%
    \begin{subfigure}[t]{0.33\textwidth}
    \begin{center}
    {\includegraphics[width = 0.99 \linewidth]{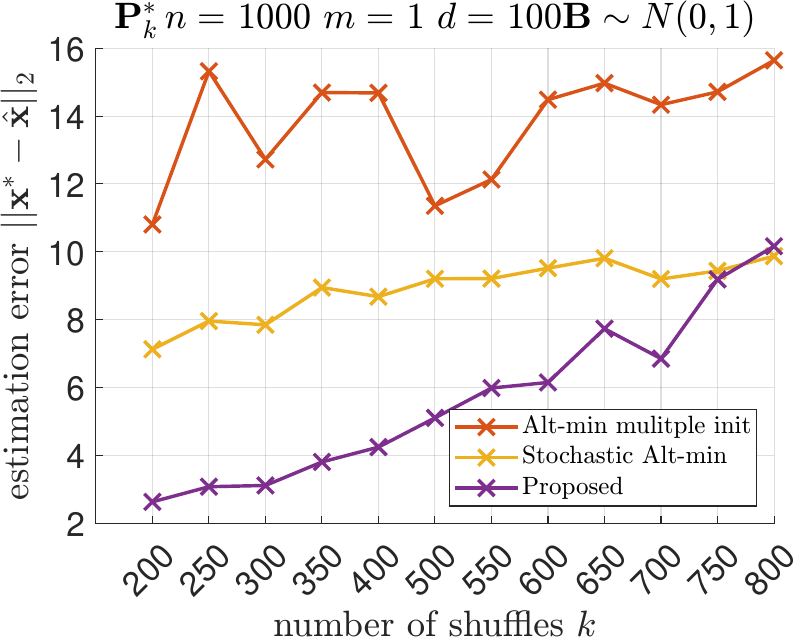}}
    \caption{}
    \label{subfig:alt-minComparison}
    \end{center}
    \end{subfigure}%
    \caption{\small{ $\Y = \P^*\B_{n \times d} \X^*_{d \times m} + \W$.  In figures (a,b,c,d), the normalized Hamming distortion $d_H/n$  is plotted on the y-axis against block size $r$ (a) and the number of shuffles (b,c,d).  Hamming distortion $d_H$ is the number of mismatches in estimate $\widehat{\P}$ of $\P^*$ and is defined as $d_H = \Sigma_{i}\mathbbm{1} (\widehat \P(i) \neq \P^*(i))$, where $\P(i)$ denotes the column index of the $1$ entry in the $i^{th}$ row of the permutation matrix $\P$. A lower value of Hamming distortion is better.} 
    }
    \label{fig:syn_sim}
\end{figure*}

We compare the proposed AltMin algorithm to several benchmark methods on both synthetic (Figure \ref{fig:syn_sim}) and real  (Table \ref{tbl:rslts}) datasets. We implemented all algorithms in MATLAB. The linear assignment problem to recover the permutation estimate $\P$ is solved by using the MATLAB \verb|matchpairs| function. The least squares estimate (line 7 of Algorithm 1 and  line 4 of Algorithm 2) is solved by computing the Moore-Penrose pseudoinverse using the MATLAB function \verb|pinv|. The convex optimization problem proposed in \cite{slawski_two_stage} is solving using CVX \cite{cvx,gb08}. The specific solver we used is SeDuMi. We also use CVX to project onto the set of doubly stochastic matrices for the benchmark method `ds+'. GitHub repository: [\url{github.com/aabbas02/ksparse-and-rlocal}].
\label{sec:results}
\begin{table*}[!]	
\centering
\hspace{0em} \renewcommand{\arraystretch}{0.95}
\begin{tabular}{|c|c|c|c|}	
    \hline & & &    \\[-2ex]	
    {\bfseries \textsf{Dataset}}  & $n$ & $d$ & $m$ \\ \hline	
    ftp     & 335  & 30 & 6     \\ \hline
    ames & 2747  &  6   & 1   \\ \hline
    scm     & 8966  & 35 & 16   \\ \hline 
scs     & 44484  &  10    & 6   \\  \hline  \hline
    air-qty & 27605  &  27    & 16  \\  \hline	
    \end{tabular}
  \caption{Description of the datasets used to compare the proposed algorithm with baseline methods. The results are given in Table \ref{tbl:rslts}. Here, $n$ is the number of data points, $d$ is the number of features, and $m$ is the number of response variables.    }	
\label{tbl:desc}	
\end{table*}
\hspace{-15.55em} 
\begin{table*}
\renewcommand{\arraystretch}{0.95}
\begin{tabular}{|c|c|c|c|c|c|c|c|c|}
    \hline & & & & & & & &    \\[-2ex]	
    {\bfseries \textsf{Dataset}}  & { $  r_{\textrm{max}} $} & $s$ & {\bfseries \textsf{Oracle}} & {\bfseries \textsf{Naive}}  & {\bfseries \textsf{RLUS}} & $\ell_2$-{\bfseries \textsf{reg}} & {\bfseries \textsf{AltMin-$k$}} & {\bfseries \textsf{AltMin-$r$}}   \\ \hline	
    ftp     & 43     & 46  & $(0.88,0)$ & $(0.44,0.70)$ & $(0.62,0.01)$   & $(0.57,0.59)$    & $(0.73,0.41)$  & $(\mathbf{0.85},\mathbf{0.17})$  \\ \hline
    ames &  311 & 61 & $(0.85,0.32)$ & $(0.37,0.75)$ & $(0.38,0.72)$  & $(0.72,0.89)$     & $(0.69,0.42)$  & $(\mathbf{0.76},\mathbf{0.32})$ \\ \hline
    scm      & 424     & 1137  & $(0.58,0)$ & $(0.21,0.79)$ & $(0.53,0.29)$   & $(0.46,0.49)$ & $(0.54,0.29)$  & $(\mathbf{0.55},\mathbf{0.25})$ \\ \hline 
    scs    & 3553    & 356 & $(0.81,0)$ & $(0.01,0.99)$ & $(0.74,0.24)$   & $(0.02,0.98)$     & $(0.73,0.33)$  & $(\mathbf{0.77},\mathbf{0.21})$ \\  \hline  \hline
    air-qty  &   95  & 366 & $(0.69,0)$ & $(0.29,0.78)$ & $(0.65,0.20)$   & $(0.64,0.21)$     & $\mathbf{(0.66,0.17)}$ & $(0.65,0.19)$ \\  \hline	
    air-qty & 46   & 744 & $(0.69,0)$ & $(0.10,0.94)$ & $(0.60,0.34)$  & $(0.26,0.79)$     & $(0.57,0.42)$  & $\mathbf{(0.62,0.32)}$ \\  \hline	
  \end{tabular}\hspace{4.65em}
  \caption{{$r_\textrm{max}$ denotes the largest block size of $\P^*_r$ and $s$ denotes the number of blocks. For a description of the datasets, see Table \ref{tbl:desc}. The methods are compared by the coefficient of determination $R_2$ and relative error $(R_2, \text{relative error})$. The relative error is $\lVert  \X^* - \widehat \X \rVert_F/\lVert \X^* \rVert_F$, where $\X^* = \B^\dagger\Y^*$ is the `oracle' regression matrix given unpermuted data $\Y^*$. The coefficient $R_2 (\widehat \X) = 1 - (\lVert \Y^* - \B \widehat \X \rVert_F/\lVert \Y^* \rVert_F)$ measures the goodness of fit for the unpermuted data. The `naive' estimate from permuted data $\Y$ is $\widehat \X = \B^\dagger \Y$.  The coefficient $R_2$ is bounded $0 \leq R_2 \leq 1$, and a higher value  of  $R_2$ indicates a better fit. }    }	
\label{tbl:rslts}	
\end{table*}
\paragraph{\textbf{Baselines.}}  We compare against six baseline methods, `$\ell_2$-regularized' \cite{slawski_two_stage}, `ds+' \cite{slawski_two_stage}, `Spectral' \cite{icml}, `Biconvex' \cite{snr}, `RLUS' \cite{ojsp},  and `Stochastic AltMin' \cite{abid2018stochastic}. The `$\ell_2$-regularized' method considers the $k$-sparse permutation model and imposes a row-wise group sparse penalty ($\ell_2$-norm regularization) on $\M$, where $\widehat \Y = \B \widehat \X + \M$. Other methods are discussed in the following paragraphs. For the proposed algorithms, \eqref{eq:P_update}, \eqref{eq:x_update} are alternated until the change in the objective value is less than $1$ percent.

\paragraph{\textbf{Synthetic data generation.}} To simulate $\Y = \P^*\B_{n \times d} \X^*_{d \times m} + \W$, we generate data by drawing the entries of matrices $\B, \X^*$ and $\W$  from the normal distribution. Subsequently, $\W$ is scaled by $\sigma$ to set $\textrm{SNR} \triangleq \lVert \X^* \rVert_F^2/(m\sigma^2) = 100$. The permutation matrices $\P^*_r$ and $\P^*_k$ are sampled uniformly from the set of $r$-local and $k$-sparse permutations, respectively.  The results are averaged over 15 Monte-Carlo runs.

\paragraph{\textbf{Enforcing block-diagonal constraints.}} To adapt the `Spectral', `$\ell_2$-regularized', and `Biconvex' methods to the $r$-local model, we add a constraint enforcing the permutation estimates to be block-diagonal.'
\begin{itemize}
\item The Spectral method in \cite{zhang2019permutation} studies the unlabeled sensing model $\Y = \P_i\B\X+\mathbf{W}$, where $\mathbf{W}$
is an additive Gaussian noise, with no assumption made on the structure of the permutation, i.e.,
the underlying permutation is generic. The proposed algorithm in \cite{zhang2019permutation} estimates $\P$ using 
\begin{equation}
\label{eq:icml_update}
\min_{\P \in \Pi_{n}}\quad  \langle \P ,  \Y\Y^\intercal \B\B^\intercal \rangle.
\end{equation}
In the case that the underlying permutation is $r$-local, first recall that $s$ denotes the number of blocks in the $r$-local permutation and $r_i$ is the size of the $i$-th block. The unlabeled sensing model reduces to 
\[
\Y_i = \P_i \B_i\X_i+\mathbf{W}_i,
\]
where $\Y_i \in \mathbb{R}^{r_i \times m}, \B_i \in \mathbb{R}^{r_i \times d}$ and $\mathbf{W}_i \in \mathbb{R}^{r_i \times m}$ respectively denote blocks of the matrices $\Y \in \mathbb{R}^{n \times m}, \B \in \mathbb{R}^{n \times d}$
and $\mathbf{W} \in \mathbb{R}^{n \times m}$. Therefore, accounting for the structure of the $r$-local permutation, we modify \eqref{eq:icml_update} as follows:
\[
\min_{\P \in \Pi_{r_i} }\quad   \langle \P_i ,  \Y_i\Y_i^\intercal \B_i\B_i^\intercal \rangle \quad \forall i \in [s],
\]
which are $s$ linear assignment problems over the sets of permutation matrices of size $r_i$.
\item For the `Biconvex' algorithm, we  modify the $\P_1,\P_2$ updates in Algorithm 1  from  \cite{snr}. Let $\mathbf{C}_1 \triangleq  -(\Y\Y^\intercal)\P_2^{(t)}\P_\B^\intercal + \mu - \rho\P_2^{(t)}$, where $\mu$ and $\rho$ are ADMM parameters, and $\P_{\B}$ is the projection matrix onto the column-span of $\B$.  Instead of updating $\P_1^{(t+1)} = \argmin_{\P \in \Pi_n} \langle \P, \mathbf{C}_1 \rangle$, we enforce the block diagonal constraint by updating $\P_{1,i}^{(t+1)} =  \argmin_{\P_i \in \Pi_{r_i}} \langle  \P_i,\mathbf{C}_{1,i}  \rangle $, where $\mathbf{C}_{1,i} \in \mathbb{R}^{r \times r}$ is the matrix comprised of entries from the $r_i$ rows and $r_i$ columns in the $i$-the block of $\mathbf{C}$. We update $\P_2^{(t+1)}$ in a similar manner. 
\item The `$\ell_2$-regularized' method considers the $k$-sparse permutation model and imposes a row-wise group sparse penalty on $\M$, where $\widehat{\Y} = \B \widehat{\X} + \M$. Specifically, the proposed estimate is obtained by solving the minimization problem:
\[
\min_{\X, \M}\quad \lVert \Y - \B \X \rVert_F^2 + \lambda \sum_{i=1}^{n} \lVert \M_i \rVert_2,
\]
where $\M_i$ denotes the $i$-th row of $\M$. We do not modify this minimization over $\X$ and $\M$. Instead, We substitute this estimate, denoted by $\widehat{\X}$, and let $\widehat{\Y} = \B \widehat{\X}$. To estimate the underlying permutation, we consider the following minimization problem $\min_{\P}\quad  \lVert \Y - \P \widehat{\Y} \rVert_F^2.$ We note the following equalities:
\begin{align*}
\argmin_{\P} \lVert \Y - \P \widehat{\Y} \rVert_F^2 
& = \argmin_{\P}\,\, \lVert \Y \rVert_F^2 + \lVert \P \widehat{\Y} \rVert_F^2 - 2 \langle \Y, \P \widehat{\Y} \rangle \\
& = \argmax_{\P}\, \langle \Y, \P \widehat{\Y} \rangle  = \argmax_{\P}\, \Tr(\Y^\top \P \widehat{\Y}) 
 = \argmax_{\P}\, \langle \Y \widehat{\Y}^\top,\P \rangle.
\end{align*}
In the case that the underlying permutation is $r$-local, the above minimization problem decouples, and we obtain a block-wise linear assignment problem:
\[
\widehat{\P}_i \gets \argmin_{\P_i \in \Pi_{r_i}} - \langle \Y_i \widehat{\Y}_i^\top, \P_i \rangle,
\]
where $\widehat{\Y}_i \in \mathbb{R}^{r_i \times m}$ and $\Y_i \in \mathbb{R}^{r_i \times m}$ denote blocks of the matrices $\widehat{\Y}$ and $\Y$, respectively.

\end{itemize}

\paragraph{\textbf{Figures \ref{subfig:rLocalSynth}, \ref{subfig:randnB}.}} 
The results show that our algorithm recovers $\P^*$ with decreasing block size $r$ and number of shuffles $k$. This confirms the conclusions of Theorems \ref{thm:multiViewfixedXrandomB},\ref{thm:multiViewRandomX}, and \ref{thm:kSparse} as the initialization improves with lower values of $r$ and $k$. The proposed AltMin algorithm is also applicable to both models and computationally scalable. For $r=125$ (Fig. \ref{subfig:rLocalSynth}) and $k=850$  (Fig. \ref{subfig:randnB}),  MATLAB runtime with 16 Gb RAM and 9-th Gen. 4-core processor is less than a second.

\paragraph{\textbf{Figure \ref{subfig:randB}.}} The entries of the measurement matrix $\B$ are sampled i.i.d. from the uniform $[0, 1]$ distribution.  Compared to the case for Gaussian $\B$ (Fig. \ref{subfig:randnB}), the performance of the `Spectral' and `RLUS' methods deteriorates significantly. This is because both algorithms consider quadratic measurements $\Y\Y^\top$. Specifically, the `Spectral' method is based on the spectral initialization technique \cite{Spectral}, which assumes a Gaussian measurement matrix. In contrast, the performance of the proposed  and  `$\ell_2$-regularized' methods does not deteriorate. 

\paragraph{\textbf{Figure \ref{subfig:ds+}.}} The `ds+' algorithm \cite{slawski_two_stage} considers the convex relaxation of \eqref{eq:obj} by minimizing the objective over the set of doubly stochastic matrices. Assuming a known upper bound on the number of shuffles $k$, `ds+'  constrains $\langle \mathbf{I},\P \rangle \geq n-k$. To project onto the set of doubly stochastic matrices, each iteration of `ds+' minimizes a linear program which greatly increases its run-time. The proposed AltMin algorithm optimizes the same objective, but over the set of permutation matrices. This results in a simpler linear assignment problem, which is why AltMin is  faster and outperforms `ds+'.

\paragraph{\textbf{Figure \ref{subfig:alt-minComparison}.}} We compare to the method in \cite{abid2018stochastic} which considers the $m=1$ single-view setup and proposes stochastic alternating minimization (S.AltMin)  to optimize \eqref{eq:obj}. S.AltMin updates $\P$ 50 times in each iteration and its run-time is 50 times that of the proposed algorithm.  \cite{abid2018stochastic} also proposes AltMin with multiple initializations for $\widehat\P^{(0)}$ with a similarly high run-time. The results show that our algorithm (AltMin with $\widehat\P^{(0)} = \mathbf{I}$ initialization) outperforms both S.AltMin and AltMin with multiple initializations. Check what AltMin refers to.

\paragraph{\textbf{TABLE \ref{tbl:rslts}.}} For the linked linear regression problem (Section \ref{subsec:linkedLinearRegression}), we report results on the three datasets from \cite{slawski_two_stage}, available at \cite{atp1d, scs,air-qty}, and the Ames \cite{amesHousing} housing datasets.  For all datasets, the columns of the response variables $\Y$ and the design matrix $\B$ are centered (zero-mean). $\B$ is also replaced by its top $d$ principal components. For the `scs' dataset, one of the 7 response variables is excluded to improve the model fit.   For `ftp', `ames', `scm', `ames' and `scs', the values of a feature (column of the design matrix) are rounded and data points with the same feature value are assigned to the same block and permuted.  

The `air-qty' dataset contains time-stamped readings with year, month, day, and hour information and we follow the experimental setup designed in the `Case Study' in Section 4 of \cite{slawski_two_stage}: in row 4 (5) of Table \ref{tbl:rslts}, readings with the same month and day (day and hour) are assigned to the same block and permuted. The regression model for `air-qty' is also as defined  in \cite{slawski_two_stage},   and we also use a moving-median filter with window size 32 to remove outliers. The proposed AltMin algorithm outperforms the competing baselines. `AltMin-$k$', i.e., AltMin initialized to $\widehat \P^{(0)} = \mathbf{I}$, is also competitive, possibly because permuting correlated rows does not greatly corrupt the design matrix $\B$. Results for `Spectral' and `Biconvex'  are  omitted because the methods were not competitive.

\section{Conclusion}
In this paper, we studied a fast alternating minimization algorithm for the unlabeled sensing problem under two structured permutation models: $k$-sparse and $r$-local. The initialization of this non-convex algorithm plays a crucial role in its performance. To address this, we proposed two initialization strategies tailored to the respective permutation models.
Our primary contribution lies in the characterization of the initialization error, providing theoretical insights into its impact on algorithm performance. Additionally, we show the competitive performance of the algorithm on both synthetic and real datasets.
While the current work focuses on analyzing the initialization, one direction for future research is to study the rate of convergence and establish conditions under which the algorithm converges provably to the unknown parameters of the unlabeled sensing problem.

\section{Acknowledgements}
Abiy Tasissa thanks Professor R. Adamczak for correspondence on \cite{adamczak2015note}. Abiy Tasissa's research is partially supported by NSF DMS 2208392. 
\bibliographystyle{IEEEtranN}
\bibliography{AltMin_ULS_initialization}

\begin{thebibliography}{50}
\providecommand{\natexlab}[1]{#1}
\providecommand{\url}[1]{#1}
\csname url@samestyle\endcsname
\providecommand{\newblock}{\relax}
\providecommand{\bibinfo}[2]{#2}
\providecommand{\BIBentrySTDinterwordspacing}{\spaceskip=0pt\relax}
\providecommand{\BIBentryALTinterwordstretchfactor}{4}
\providecommand{\BIBentryALTinterwordspacing}{\spaceskip=\fontdimen2\font plus
\BIBentryALTinterwordstretchfactor\fontdimen3\font minus
  \fontdimen4\font\relax}
\providecommand{\BIBforeignlanguage}[2]{{%
\expandafter\ifx\csname l@#1\endcsname\relax
\typeout{** WARNING: IEEEtranN.bst: No hyphenation pattern has been}%
\typeout{** loaded for the language `#1'. Using the pattern for}%
\typeout{** the default language instead.}%
\else
\language=\csname l@#1\endcsname
\fi
#2}}
\providecommand{\BIBdecl}{\relax}
\BIBdecl

\bibitem[Zhang et~al.(2019)Zhang, Slawski, and Li]{snr}
H.~Zhang, M.~Slawski, and P.~Li, ``Permutation recovery from multiple
  measurement vectors in unlabeled sensing,'' in \emph{2019 IEEE International
  Symposium on Information Theory (ISIT)}, 2019, pp. 1857--1861.

\bibitem[Slawski et~al.(2020{\natexlab{a}})Slawski, Ben-David, and
  Li]{slawski_two_stage}
M.~Slawski, E.~Ben-David, and P.~Li, ``Two-stage approach to multivariate
  linear regression with sparsely mismatched data.'' \emph{J. Mach. Learn.
  Res.}, vol.~21, no. 204, pp. 1--42, 2020.

\bibitem[Zhang and Li(2020{\natexlab{a}})]{zhang2019permutation}
\BIBentryALTinterwordspacing
H.~Zhang and P.~Li, ``Optimal estimator for unlabeled linear regression,'' in
  \emph{Proceedings of the 37th International Conference on Machine Learning},
  ser. Proceedings of Machine Learning Research, H.~D. III and A.~Singh, Eds.,
  vol. 119.\hskip 1em plus 0.5em minus 0.4em\relax PMLR, 13--18 Jul 2020, pp.
  11\,153--11\,162. [Online]. Available:
  \url{https://proceedings.mlr.press/v119/zhang20n.html}
\BIBentrySTDinterwordspacing

\bibitem[Slawski et~al.(2020{\natexlab{b}})Slawski, Rahmani, and
  Li]{slawski2020sparse}
M.~Slawski, M.~Rahmani, and P.~Li, ``A sparse representation-based approach to
  linear regression with partially shuffled labels,'' in \emph{Uncertainty in
  Artificial Intelligence}.\hskip 1em plus 0.5em minus 0.4em\relax PMLR, 2020,
  pp. 38--48.

\bibitem[Shi et~al.(2021)Shi, Li, and Cai]{spherical}
X.~Shi, X.~Li, and T.~Cai, ``Spherical regression under mismatch corruption
  with application to automated knowledge translation,'' \emph{Journal of the
  American Statistical Association}, vol. 116, no. 536, pp. 1953--1964, 2021.

\bibitem[Slawski and Ben-David(2019)]{slawski2019linear}
M.~Slawski and E.~Ben-David, ``Linear regression with sparsely permuted data,''
  \emph{Electronic Journal of Statistics}, vol.~13, no.~1, pp. 1--36, 2019.

\bibitem[Abbasi et~al.(2021)Abbasi, Tasissa, and Aeron]{ojsp}
A.~A. Abbasi, A.~Tasissa, and S.~Aeron, ``R-local unlabeled sensing: A novel
  graph matching approach for multiview unlabeled sensing under local
  permutations,'' \emph{IEEE Open Journal of Signal Process.}, vol.~2, pp.
  309--317, 2021.

\bibitem[Li et~al.(2022)Li, Aubry, De~Maio, Fu, and Marano]{marano}
D.~Li, A.~Aubry, A.~De~Maio, Y.~Fu, and S.~Marano, ``Detection by block- and
  band-permuted data,'' \emph{IEEE Transactions on Signal Processing}, vol.~70,
  pp. 5778--5790, 2022.

\bibitem[A.~Abbasi et~al.(2022)A.~Abbasi, Tasissa, and Aeron]{icassp_r_local}
A.~A.~Abbasi, A.~Tasissa, and S.~Aeron, ``r-local unlabeled sensing: Improved
  algorithm and applications,'' in \emph{ICASSP 2022-2022 IEEE International
  Conference on Acoustics, Speech and Signal Processing (ICASSP)}.\hskip 1em
  plus 0.5em minus 0.4em\relax IEEE, 2022, pp. 5593--5597.

\bibitem[Wang et~al.(2023)Wang, Ben-David, and Slawski]{wang2023regularization}
Z.~Wang, E.~Ben-David, and M.~Slawski, ``Regularization for shuffled data
  problems via exponential family priors on the permutation group,'' in
  \emph{International Conference on Artificial Intelligence and
  Statistics}.\hskip 1em plus 0.5em minus 0.4em\relax PMLR, 2023, pp.
  2939--2959.

\bibitem[Unnikrishnan et~al.(2018)Unnikrishnan, Haghighatshoar, and
  Vetterli]{unnikrishnan2018unlabeled}
J.~Unnikrishnan, S.~Haghighatshoar, and M.~Vetterli, ``Unlabeled sensing with
  random linear measurements,'' \emph{IEEE Trans. Inf. Theory}, vol.~64, no.~5,
  pp. 3237--3253, 2018.

\bibitem[{Dokmanić}(2019)]{Dokmanic}
I.~{Dokmanić}, ``Permutations unlabeled beyond sampling unknown,'' \emph{IEEE
  Signal Processing Letters}, vol.~26, no.~6, pp. 823--827, 2019.

\bibitem[{Pananjady} et~al.(2018){Pananjady}, {Wainwright}, and
  {Courtade}]{pananjady}
A.~{Pananjady}, M.~J. {Wainwright}, and T.~A. {Courtade}, ``Linear regression
  with shuffled data: Statistical and computational limits of permutation
  recovery,'' \emph{IEEE Trans. Inf. Theory}, vol.~64, no.~5, pp. 3286--3300,
  2018.

\bibitem[{Emiya} et~al.(2014){Emiya}, {Bonnefoy}, {Daudet}, and
  {Gribonval}]{emiya}
V.~{Emiya}, A.~{Bonnefoy}, L.~{Daudet}, and R.~{Gribonval}, ``Compressed
  sensing with unknown sensor permutation,'' in \emph{2014 IEEE Int. Conference
  on Acoustics, Speech and Signal Processing (ICASSP)}, 2014, pp. 1040--1044.

\bibitem[Peng and Tsakiris(2020)]{concave}
L.~Peng and M.~C. Tsakiris, ``Linear regression without correspondences via
  concave minimization,'' \emph{IEEE Signal Proces. Letters}, vol.~27, pp.
  1580--1584, 2020.

\bibitem[Peng et~al.(2019)Peng, Song, Tsakiris, Choi, Kneip, and
  Shi]{header_free}
L.~Peng, X.~Song, M.~C. Tsakiris, H.~Choi, L.~Kneip, and Y.~Shi,
  ``Algebraically-initialized expectation maximization for header-free
  communication,'' in \emph{IEEE Int. Conf. on Acous., Speech and Signal
  Process. (ICASSP)}.\hskip 1em plus 0.5em minus 0.4em\relax IEEE, 2019, pp.
  5182--5186.

\bibitem[Abid and Zou(2018)]{abid2018stochastic}
A.~Abid and J.~Zou, ``A stochastic expectation-maximization approach to
  shuffled linear regression,'' in \emph{2018 56th Annual Allerton Conference
  on Communication, Control, and Computing (Allerton)}.\hskip 1em plus 0.5em
  minus 0.4em\relax IEEE, 2018, pp. 470--477.

\bibitem[{Pananjady} et~al.(2017){Pananjady}, {Wainwright}, and
  {Courtade}]{Levsort}
A.~{Pananjady}, M.~J. {Wainwright}, and T.~A. {Courtade}, ``Denoising linear
  models with permuted data,'' in \emph{2017 IEEE Int. Symposium on Inf. Theory
  (ISIT)}, 2017, pp. 446--450.

\bibitem[Zhang and Li(2020{\natexlab{b}})]{icml}
H.~Zhang and P.~Li, ``Optimal estimator for unlabeled linear regression,'' in
  \emph{Int. Conference on Machine Learning}.\hskip 1em plus 0.5em minus
  0.4em\relax PMLR, 2020, pp. 11\,153--11\,162.

\bibitem[Wang et~al.(2018)Wang, Zhu, Blum, Willett, Marano, Matta, and
  Braca]{wang2018signal}
G.~Wang, J.~Zhu, R.~S. Blum, P.~Willett, S.~Marano, V.~Matta, and P.~Braca,
  ``Signal amplitude estimation and detection from unlabeled binary quantized
  samples,'' \emph{IEEE Transactions on Signal Processing}, vol.~66, no.~16,
  pp. 4291--4303, 2018.

\bibitem[Haghighatshoar and Caire(2017)]{haghighatshoar2017signal}
S.~Haghighatshoar and G.~Caire, ``Signal recovery from unlabeled samples,''
  \emph{IEEE Transactions on Signal Processing}, vol.~66, no.~5, pp.
  1242--1257, 2017.

\bibitem[Liu and Zhu(2018)]{liu2018signal}
Z.~Liu and J.~Zhu, ``Signal detection from unlabeled ordered samples,''
  \emph{IEEE Communications Letters}, vol.~22, no.~12, pp. 2431--2434, 2018.

\bibitem[Marano and Willett(2019)]{marano2019algorithms}
S.~Marano and P.~K. Willett, ``Algorithms and fundamental limits for unlabeled
  detection using types,'' \emph{IEEE Transactions on Signal Processing},
  vol.~67, no.~8, pp. 2022--2035, 2019.

\bibitem[Marano and Willett(2020)]{marano2020making}
S.~Marano and P.~Willett, ``Making decisions by unlabeled bits,'' \emph{IEEE
  Transactions on Signal Processing}, vol.~68, pp. 2935--2947, 2020.

\bibitem[Sutton et~al.(2023)Sutton, Willett, Marano, and
  Bar-Shalom]{sutton2023identity}
Z.~Sutton, P.~Willett, S.~Marano, and Y.~Bar-Shalom, ``Identity-aware decision
  network communication budgeting: Is who as important as what?'' \emph{IEEE
  Transactions on Aerospace and Electronic Systems}, vol.~59, no.~5, pp.
  5203--5217, 2023.

\bibitem[Sun and Zou(2023)]{sun2023quickest}
Z.~Sun and S.~Zou, ``Quickest anomaly detection in sensor networks with
  unlabeled samples,'' \emph{IEEE Transactions on Signal Processing}, vol.~71,
  pp. 873--887, 2023.

\bibitem[Murray(2016)]{murray2016probabilistic}
J.~S. Murray, ``Probabilistic record linkage and deduplication after indexing,
  blocking, and filtering,'' \emph{Journal of Privacy and Confidentiality 7
  (1).}, 2016.

\bibitem[Lahiri and Larsen(2005)]{lahiri2005regression}
P.~Lahiri and M.~D. Larsen, ``Regression analysis with linked data,''
  \emph{Journal of the American statistical association}, vol. 100, no. 469,
  pp. 222--230, 2005.

\bibitem[Han and Lahiri(2019)]{han2019statistical}
Y.~Han and P.~Lahiri, ``Statistical analysis with linked data,''
  \emph{International Statistical Review}, vol.~87, pp. S139--S157, 2019.

\bibitem[Kim and Chambers(2012)]{kim2012regression}
G.~Kim and R.~Chambers, ``Regression analysis under incomplete linkage,''
  \emph{Computational Statistics \& Data Analysis}, vol.~56, no.~9, pp.
  2756--2770, 2012.

\bibitem[Wang et~al.(2022)Wang, Ben-David, Diao, and
  Slawski]{wang2022regression}
Z.~Wang, E.~Ben-David, G.~Diao, and M.~Slawski, ``Regression with linked
  datasets subject to linkage error,'' \emph{Wiley Interdisciplinary Reviews:
  Computational Statistics}, vol.~14, no.~4, p. e1570, 2022.

\bibitem[Wang et~al.(2020)Wang, Marano, Zhu, and Xu]{wang2020target}
G.~Wang, S.~Marano, J.~Zhu, and Z.~Xu, ``Target localization by unlabeled range
  measurements,'' \emph{IEEE Transactions on Signal Processing}, vol.~68, pp.
  6607--6620, 2020.

\bibitem[Koka et~al.(2024)Koka, Tsakiris, Muma, and Haro]{koka2024shuffled}
T.~Koka, M.~C. Tsakiris, M.~Muma, and B.~B. Haro, ``Shuffled multi-channel
  sparse signal recovery,'' \emph{Signal Processing}, p. 109579, 2024.

\bibitem[Ravikumar and Cohen(2012)]{graph_rl}
P.~Ravikumar and W.~Cohen, ``A hierarchical graphical model for record
  linkage,'' \emph{arXiv preprint arXiv:1207.4180}, 2012.

\bibitem[Hsu et~al.(2012)Hsu, Kakade, and Zhang]{hsu2012tail}
D.~Hsu, S.~Kakade, and T.~Zhang, ``A tail inequality for quadratic forms of
  subgaussian random vectors,'' \emph{Electronic Communications in
  Probability}, vol.~17, pp. 1--6, 2012.

\bibitem[Vershynin(2018)]{vershynin2018high}
R.~Vershynin, \emph{High-dimensional probability: An introduction with
  applications in data science}.\hskip 1em plus 0.5em minus 0.4em\relax
  Cambridge university press, 2018, vol.~47.

\bibitem[Laurent and Massart(2000)]{10.1214/aos/Laurent}
\BIBentryALTinterwordspacing
B.~Laurent and P.~Massart, ``{Adaptive estimation of a quadratic functional by
  model selection},'' \emph{The Annals of Statistics}, vol.~28, no.~5, pp. 1302
  -- 1338, 2000. [Online]. Available:
  \url{https://doi.org/10.1214/aos/1015957395}
\BIBentrySTDinterwordspacing

\bibitem[Woodruff et~al.(2014)]{woodruff2014sketching}
D.~P. Woodruff \emph{et~al.}, ``Sketching as a tool for numerical linear
  algebra,'' \emph{Foundations and Trends{\textregistered} in Theoretical
  Computer Science}, vol.~10, no. 1--2, pp. 1--157, 2014.

\bibitem[Duxbury et~al.(2016)Duxbury, Granlund, Gujarathi, Juhas, and
  Billinge]{duxbury2016unassigned}
P.~M. Duxbury, L.~Granlund, S.~Gujarathi, P.~Juhas, and S.~J. Billinge, ``The
  unassigned distance geometry problem,'' \emph{Discrete Applied Mathematics},
  vol. 204, pp. 117--132, 2016.

\bibitem[Duxbury et~al.(2022)Duxbury, Lavor, Liberti, and
  de~Salles-Neto]{duxbury2022unassigned}
P.~Duxbury, C.~Lavor, L.~Liberti, and L.~L. de~Salles-Neto, ``Unassigned
  distance geometry and molecular conformation problems,'' \emph{Journal of
  Global Optimization}, pp. 1--10, 2022.

\bibitem[Skiena et~al.(1990)Skiena, Smith, and Lemke]{Lemke2003}
S.~S. Skiena, W.~D. Smith, and P.~Lemke, ``Reconstructing sets from interpoint
  distances,'' in \emph{Proceedings of the sixth annual symposium on
  Computational geometry}, 1990, pp. 332--339.

\bibitem[Rudelson and Vershynin(2009)]{rudelson2009smallest}
M.~Rudelson and R.~Vershynin, ``Smallest singular value of a random rectangular
  matrix,'' \emph{Communications on Pure and Applied Mathematics: A Journal
  Issued by the Courant Institute of Mathematical Sciences}, vol.~62, no.~12,
  pp. 1707--1739, 2009.

\bibitem[Grant and Boyd(2014)]{cvx}
M.~Grant and S.~Boyd, ``{CVX}: Matlab software for disciplined convex
  programming, version 2.1,'' \url{http://cvxr.com/cvx}, Mar. 2014.

\bibitem[Grant and Boyd(2008)]{gb08}
------, ``Graph implementations for nonsmooth convex programs,'' in
  \emph{Recent Advances in Learning and Control}, ser. Lecture Notes in Control
  and Information Sciences, V.~Blondel, S.~Boyd, and H.~Kimura, Eds.\hskip 1em
  plus 0.5em minus 0.4em\relax Springer-Verlag Limited, 2008, pp. 95--110,
  \url{http://stanford.edu/~boyd/graph_dcp.html}.

\bibitem[Luo et~al.(2019)Luo, Alghamdi, and Lu]{Spectral}
W.~Luo, W.~Alghamdi, and Y.~M. Lu, ``Optimal spectral initialization for signal
  recovery with applications to phase retrieval,'' \emph{IEEE Transactions on
  Signal Processing}, vol.~67, no.~9, pp. 2347--2356, 2019.

\bibitem[Spyromitros-Xioufis et~al.(2016)Spyromitros-Xioufis, Tsoumakas,
  Groves, and Vlahavas]{atp1d}
E.~Spyromitros-Xioufis, G.~Tsoumakas, W.~Groves, and I.~Vlahavas,
  ``Multi-target regression via input space expansion: treating targets as
  inputs,'' \emph{Machine Learning}, vol. 104, no.~1, pp. 55--98, 2016.

\bibitem[Rasmussen and Williams(2006)]{scs}
C.~E. Rasmussen and C.~K.~I. Williams, \emph{Gaussian Processes for Machine
  Learning}.\hskip 1em plus 0.5em minus 0.4em\relax Cambridge, MA: MIT Press,
  2006, dataset available at \url{http://www.gaussianprocess.org/gpml/data/}.

\bibitem[Chen(2017)]{air-qty}
S.~Chen, ``{Beijing Multi-Site Air Quality},'' UCI Machine Learning Repository,
  2017, {DOI}: https://doi.org/10.24432/C5RK5G.

\bibitem[De~Cock(2011)]{amesHousing}
D.~De~Cock, ``Ames, iowa: Alternative to the boston housing data as an end of
  semester regression project,'' \emph{Journal of Statistics Education},
  vol.~19, no.~3, 2011.

\bibitem[Adamczak(2015)]{adamczak2015note}
R.~Adamczak, ``A note on the {Hanson-Wright} inequality for random vectors with
  dependencies,'' \emph{Electronic Communications in Probability}, vol.~20, pp.
  1--13, 2015.

\end{thebibliography}
\appendix
\label{sec:AppndxA}
\section{Proof Of Theorem 
\ref{thm:multiViewRandomX}}

\begin{proof}
Recall that $\tilde \x_{\text{err}}$ defined in \eqref{eq:lower_tail},  \eqref{eq:upper_tail} is sub-exponential (see Lemma \ref{lem:subExp}). For $t \geq 0$, 
\begin{align}
    \Pr[\lvert \tilde \x_{\text{err}} - \mathbb{E}[\tilde \x_{\text{err}}] \rvert \geq t] &= \Pr[\lvert \tilde \x_{\text{err}} - C_{d-s} - \mathbb{E}[\tilde \x_{\text{err}} - C_{d-s}] \rvert \geq t] \nonumber\\ 
    &\leq 2\exp(-c't/\lVert \tilde \x_{\text{err}} - C_{d-s} - \mathbb{E}[\tilde \x_{\text{err}} - C_{d-s}]\rVert_{\varphi_1}) \label{eq:subExpBound-1}\\
   & \leq 2 \exp (-c't/K_{d-s}),         \label{eq:subExpBound}
\end{align}
where \eqref{eq:subExpBound-1}  follows from noting that $\tilde \x_{\text{err}} - C_{d-s}$ is sub-exponential,  shown in Lemma \ref{lem:subExp}. \eqref{eq:subExpBound} is by the centering property, (see proposition 2.7.1 (a), (d) in \cite{vershynin2018high}), which bounds 
$$ \lV \tilde \x_{\text{err}} - C_{d-s} - \mathbb{E}[\tilde \x_{\text{err}} - C_{d-s}] \rV_{\varphi_1} \leq C \lV \tilde \x_{\text{err}} - C_{d-s} \rV_{\varphi_1},$$ and that $\lV \tilde \x_{\text{err}} - C_{d-s} \rV_{\varphi_1} = K_{d-s}$, by Lemma \ref{lem:subExp}. For $z \geq 0$, $\lvert z - 1 \rvert \geq \delta  \implies \lvert z^2 - 1 \rvert \geq \max(\delta,\delta^2)$. Therefore,

\begin{align}
     \Pr\Big[\Big\lvert \frac{ \sqrt{\tilde \x_{\text{err}}} - \sqrt{\mathbb{E}[\tilde \x_{\text{err}}]}}{\sqrt{\mathbb{E}[\tilde \x_{\text{err}}]}} \Big \rvert \geq \delta\Big] & 
    \leq \Pr \Big[\Big\lvert \frac{ \tilde \x_{\text{err}} - \mathbb{E}[\tilde \x_{\text{err}}]}{\mathbb{E}[\tilde \x_{\text{err}}]} \Big \rvert \geq \delta^2 \Big] \nonumber \\ 
   &  \hspace{-2.5em} \leq 2 \exp(-c'\mathbb{E}[\tilde \x_{\text{err}}]\delta^2/K_{d-s}), \label{eq:sqrtBound0}
\end{align}
where \eqref{eq:sqrtBound0} follows by substituting $t = \delta^2$ in \eqref{eq:subExpBound}. We make a change of variable  $t = \delta \sqrt{\mathbb{E}[\tilde \x_{\text{err}}]}$ in \eqref{eq:sqrtBound0} and obtain
\begin{align}
     \Pr\big[\big\lvert  \sqrt{\tilde \x_{\text{err}}} - \sqrt{\mathbb{E}[\tilde \x_{\text{err}}]} \big \rvert \geq t \big] &\leq 2 \exp(-c't^2/K_{d-s}).
    \label{eq:sqrtBound1}
\end{align}
From \eqref{eq:sqrtBound1} and proposition 2.5.2 (i), (iv) and definition 2.5.6 in \cite{vershynin2018high}, the sub-Gaussian norm squared is bounded as
\begin{equation}
\lVert \sqrt{\tilde \x_{\text{err}}} - \sqrt{\mathbb{E}[\tilde \x_{\text{err}}]} \rVert_{\varphi_2}^2 = c'K_{d-s}. \label{eq:subgNrmBnd}\end{equation}
Applying Hoeffding's inequality \eqref{eq:Hoeffding} to $\sum \sqrt{\tilde \x_{\text{err}, j}} - \sqrt{\mathbb{E}[\tilde \x_{\text{err}, j}]}$ and  substituting $\mathbb{E}[\sqrt{\tilde \x_{\text{err},j} }] - \sqrt{\mathbb{E}[\tilde \x_{\text{err},j} ]} \leq 0$ (Jensen's),  
\begin{align}
    \hspace{-1em} \Pr \big[ \sum_{j=1}^{j=m} \sqrt{\tilde \x_{\text{err},j}} - \sqrt{\mathbb{E}[\tilde \x_{\text{err},j}]} \geq t \big] \leq 2 \exp \left(\frac{-c't^2}{mK_{d-s}}\right) .\label{eq:HfdngBound}
\end{align}
For non-negative random variable $X$, 
$ \mathbb{E}[X] = \int_0^\infty \Pr[X > t]dt. \label{eq:integralIdentity} $
Applied to $({\tilde \x_{\text{err}} - C_{d-s}})/{2K^2(\sqrt{d-s}+1)}$,
\begin{align}
    \mathbb{E}\Big[\frac{\tilde \x_{\text{err}} - C_{d-s}}{2K^2(\sqrt{d-s}+1)}\Big]  \leq  \int_{0}^{\infty} \exp(-t) dt  = 1. \label{eq:subtuteTailBound}
\end{align}
In \eqref{eq:subtuteTailBound}, we have substituted the tail probability upper bounds from Lemma \ref{lem:subExp}. Substituting $C_{d-s}$ from \eqref{eq:random_x} in \eqref{eq:subtuteTailBound},
\begin{equation}	
         \mathbb{E}[\tilde \x_{\text{err}}] \leq K^2(d-s + \frac{5}{2}\sqrt{d-s} + 2).	
         \label{eq:expctnUpprBnd}
\end{equation}
By definition in \eqref{eq:lower_tail}, \eqref{eq:upper_tail},  $ \lVert \x^*_j - \hat \x_j  \rVert_2  \leq \sqrt{\tilde \x_{\text{err},j}}  \, \forall \, j \in [m]$. Substituting \eqref{eq:expctnUpprBnd} and using $\lVert \x^*_j - \hat \x_j \rVert_2 \leq \sqrt{\tilde \x_{\text{err}}}$ in \eqref{eq:HfdngBound} completes the proof.
\end{proof}
\section{Proof of Theorem \ref{thm:kSparse}}
We first provide a high-level road map for the proof. For simplicity of explanation, we focus on the single-view problem. In that case, let $\y^* = \B\x^*$ and $\hat \y^{(0)}  =  \P^*_k \B \x^*$. Here, $\y^*$ represents the original measurement, and $\hat \y^{(0)}$ corresponds to the measurement after being shuffled by a $k$-sparse permutation. The starting point is to consider the quantity $\lVert \y^* - \hat \y^{(0)} \rVert^2$. Since this quantity is invariant under permutation, we assume, without loss of generality, that the first $k$ rows of $\B$ are shuffled. Lemma \ref{lemma:fwd} shows that $\frac{\lVert \y^* - \hat \y^{(0)} \rVert^2}{\lVert \x^* \rVert_2^2}$ can be expressed as the difference of two Chi-square random variables. Using a tail inequality for Chi-square distributed random variables (see Lemma \ref{lem:subExp}), Lemma \ref{lemma:fwd} provides a bound for this quantity. With this result established, the main crux of the proof of Theorem \ref{thm:kSparse} is to relate $\sum_{j=1}^{m} \lVert \y^*_j - \hat \y^{(0)}_j \rVert_2^2$ to $\sigma^2_{\textrm{min}}(\B) \lVert \X^* - \widehat \X^{(1)} \rVert_F^2$. 
\begin{proof} 
Let $\E$ denote the error term such that  
\begin{equation}
    \widehat \Y^{(0)} = \B \widehat \X^{(1)} + \E,
    \label{eq:orthErr}
\end{equation}
where $\widehat \X^{(1)} = \B^\dagger \widehat \Y^{(0)}$, $\widehat \Y^{(0)} = \Y = \P^*_k\B \X^*$ and $\E \perp \mathcal{R}(\B)$ is orthogonal to the range space of $\B$ since $\widehat \X^{(1)} = \underset{\X}{\argmin} \,\lVert \widehat \Y^{(0)} - \B \X \rVert_F$. Combining $\Y^* = \B\X^*$ and \eqref{eq:orthErr}, we obtain:
\begin{align}
 \lVert \Y^* - \widehat \Y^{(0)} \rVert_F^2 &= \lVert \B (\X^* - \widehat \X^{(1)}) - \E \rVert_F^2 \notag\\
 &= \lVert \B (\X^* - \widehat \X^{(1)}) \rVert_F^2  + \lV \E \rV_F^2 \notag\
 &\geq \sigma_{\min}^2(\B)\lV \X^* - \widehat \X^{(1)} \rV_F^2. \label{eq:unnYbnd}
 \end{align}
The inequality in \eqref{eq:unnYbnd} can equivalently be rewritten as:
\begin{equation}
     \sigma^2_{\textrm{min}}(\B) \lVert \X^* - \widehat \X^{(1)} \rVert_F^2 \leq \sum_{j=1}^{j=m} \lVert \y^*_j - \hat \y^{(0)}_j \rVert_2^2.
     \label{eq:result-1}
\end{equation}
Let $\mathcal{E}_j$ denote the event 
$\mathcal{E}_{j} = \{\x^*_j \mid \lVert \hat \y^*_j - \hat \y^{(0)}_j \rVert_2^2 \geq \big(2k + C\sqrt{kt} + 10t\big) \lVert \x^*_j \rVert_2^2 \} \, \forall \, j \in [m].  $
From \eqref{eq:result-1} and applying the union bound to $\cup_{j=1}^{j =m} \mathcal{E}_j$ using \eqref{eq:lemmaFwd},
\begin{equation}
    \Pr\bigg[\sigma^2_{\min} \frac{\lVert \X^* - \widehat \X^{(1)} \rVert_F^2}{\lVert \X^* \rVert_F^2} \geq 2k + C\sqrt{kt} + 10t   \bigg] \leq 7me^{-t}.
\end{equation}
For $t \geq 2\log m$,
\begin{equation}
    \Pr\bigg[ \frac{\lVert \X^* - \widehat \X^{(1)} \rVert_F^2}{\lVert \X^* \rVert_F^2} - \frac{2k}{\sigma^2_{\min}} \geq  \frac{C\sqrt{k}}{\sigma^2_{\min}}t  \bigg] \leq 7e^{-t/2}.
\end{equation}
By Theorem 1.1 in \cite{rudelson2009smallest}, for a ``tall'' Gaussian matrix $\B \in \mathbb{R}^{n \times d}$, where $\B$ is considered tall if the aspect ratio $\lambda = d/n$ satisfies $\lambda < \lambda_0$ for some sufficiently small constant $\lambda_0 > 0$, we have $\Pr[\sigma_{\min}(\B) \leq c \sqrt{n}] \leq e^{-c n}$ (see Subsection 1.2 of \cite{rudelson2009smallest}). We consider the union bound of the failure probabilities $e^{-cn}$ and $7e^{-t/2}$. A minor calculation yields that $e^{-cn}+7e^{-t/2}\le 8e^{-t/2}$ for $t\le cn$. Consequently, for $2\log(m)\le t\le cn$, we obtain
\begin{equation}
    \Pr\bigg[ \frac{\lVert \X^* - \widehat \X^{(1)} \rVert_F^2}{\lVert \X^* \rVert_F^2} - \frac{2}{c^2}\frac{k}{n} \geq  \frac{C\sqrt{k}}{n}t  \bigg] \leq 8e^{-t/2}.
\end{equation}
We now make a change of variables $t'=\frac{C\sqrt{k}}{n}t$. For $\frac{2C\sqrt{k}\log(m)}{n}\le t'\le Cc\sqrt{k}$, we have
\begin{equation}
    \Pr\bigg[ \frac{\lVert \X^* - \widehat \X^{(1)} \rVert_F^2}{\lVert \X^* \rVert_F^2} - \frac{2}{c^2}\frac{k}{n} \geq  t'  \bigg] \leq 8\exp\left(-
    \frac{t'n}{2C\sqrt{k}}\right).
\end{equation}
We make a further change of variable with $t'=\epsilon \frac{k}{n}$. For $\frac{2C\log(m)}{\sqrt{k}}\le \epsilon\le \frac{Ccn}{\sqrt{k}}$, we obtain
\begin{equation}
    \Pr\bigg[ \frac{\lVert \X^* - \widehat \X^{(1)} \rVert_F^2}{\lVert \X^* \rVert_F^2}  \geq  \left(\frac{2}{c^2}+\epsilon\right)\frac{k}{n}  \bigg] \leq 8\exp\left(-
    \frac{\epsilon \sqrt{k}}{2C}\right).
    \label{eq:relErrBnd}
\end{equation}
\eqref{eq:relErrBnd} provides a useful ($<1$) relative error bound when $k$ grows slowly with $n$, for example $k = n^{\beta}$, $\beta < 1$.
\end{proof}
\begin{lemma}
Let $\P^*_k$ be the fixed unknown $k$-sparse permutation matrix with $\langle \mathbf{I},\P^*_k \rangle = n-k$. Consider the exact unlabeled sensing problem. Let $\x^*$ be the fixed unknown vector, $\y^* = \B\x^*$,  $\hat \y^{(0)}  =  \P^*_k \B \x^*$. Assuming Gaussian $\B$,  for $k \in [n]$  and $t \geq 0$, we have
\begin{equation}     
\Pr\big[\lVert \y^* - \hat \y^{(0)} \rVert_2^2 \geq \big(2k +  4(1+\sqrt{3})\sqrt{kt} + 10t \big) \lVert \x^* \rVert_2^2 \big]  \leq 7e^{-t}. \label{eq:lemmaFwd}
\end{equation} 
\label{lemma:fwd}
\end{lemma}
\begin{proof} Under the $k$-sparse assumption on $\P^*$, the known vector $\y$ has $k$ shuffled entries. Assuming, without loss of generality, that the first $k$ rows of $\B^*$ are shuffled, 
\begin{align} 
\frac{\lVert \y^* - \hat \y^{(0)} \rVert_2^2}{\lVert \x^* \rVert_2^2} &= \frac{1}{\lVert \x^* \rVert_2^2}\sum_{i=1}^{i=k}(\b_i^\top \x^* -\b^\top_{\P(i)} \x^*)^2 \notag\\
& = 2\underbrace{\sum_{i=1}^{i=k}\frac{( \b^\top_i\x^*)^2}{\lVert \x^* \rVert_2^2}}_{\triangleq T_1} - 2\underbrace{\sum_{i=1}^{i=k}\frac{\b^\top_i \x^* \b^\top_{\P(i)}\x^*}{\lVert \x^* \rVert_2^2}}_{\triangleq T_2}
\label{eq:decompose}
\end{align}
$T_1$, defined in \eqref{eq:decompose}, is the sum of $k$ independent Chi-square random variables and is bounded, using \eqref{eq:chiSquareLower}, as follows:
\begin{equation}
    \Pr\big[T_1 \geq k + 2\sqrt{kt} + 2t\big]  \leq e^{-t}. 
    \label{eq:T1Bound}
\end{equation}
The product random variables in $T_2$ are distributed as the difference of two independent $\chi^2$ random variables. 
\begin{equation}
   \frac{\b_i^\top \x^* \b_{\P(i)}^\top \x^*  }{\lVert \x^* \rVert^2_2}   \sim \frac{1}{2} Z^1_i -\frac{1}{2} Z^2_i
   \,\,\,\,\, \forall \, i \in n-k+1, \cdots, n.
   \label{eq:diff}
\end{equation}
The random variables (rv) in \eqref{eq:diff} are not mutually independent, but each rv depends on, at most, two other rvs. To see this, let permutation $\P$ such that $\P(i) \mapsto j$, then  $\b_i^\top\x^*\b^\top_j\x^*$ is not independent of 
\begin{equation}
\b^\top_j\x^*\b_{\P(j)}^\top\x^*, \enskip  \b^\top_{\P^\top(i)}\x^*\b^\top_{i}\x^*. \label{eq:twoDepRV}
\end{equation}
The $k$ rvs in \eqref{eq:diff} can therefore be partitioned into three sets $P,Q,R$ such that the rvs within each set are independent. 
Let $k_1$ be the number of rvs in set $P$. The sum $T_P$, where 
\begin{equation}
    T_P \triangleq  \frac{1}{\lVert \x^* \rVert_2^2} \sum_{i \in P}  \b_i^\top \x^* \b_{\P(i)}^\top \x^* = \frac{1}{2}\sum_{i \in P} Z^1_i - \frac{1}{2}\sum_{i \in P} Z^2_i,
    \label{eq:setPDiff}
\end{equation}
is upper bounded in probability as
\begin{equation}
    \Pr[T_P \leq -2 \sqrt{k_1 t} - t ] \leq 2\exp(-t). \label{eq:TPBound}
\end{equation}
\eqref{eq:TPBound} follows from applying the union bound to  probabilities $p_1,p_2$ which are given by
\begin{align}
    p_1 &= \Pr\big [\sum_{i \in P} Z^1_i  \leq k_1 - 2\sqrt{k_1 t}  \big ] \leq e^{-t} \label{eq:unnBound0}, \\
    p_2 &= \Pr\big [\sum_{i \in P} Z^2_i  \geq k_1 + 2 \sqrt{k_1 t} + 2t \big ] \leq e^{-t} \label{eq:unnBound1},
\end{align}
and bounding $p_1,p_2$ using the tail inequalities  \eqref{eq:chiSquareLower} and \eqref{eq:chiSquareUpper}, respectively. Defining $T_Q,T_R$ similarly to \eqref{eq:setPDiff}, with cardinalities $k_2$, $k_3$ and applying the union bound as in \eqref{eq:unnBound0}, \eqref{eq:unnBound1} gives 
\begin{equation} \Pr\big[T_2 \leq - 2(\sqrt{k_1 t} + \sqrt{k_2 t} +  \sqrt{k_3 t}) - 3t \big] \leq 6e^{-t}, \label{eq:T2bnd} \end{equation}
where $T_2  = T_P+T_Q+T_R$. Since $\lVert [\sqrt{k_1} \enskip \sqrt{k_2} \enskip \sqrt{k_3} \ ]^\top \rVert_1^2 \leq 3 \lVert \cdot \rVert_2^2  =  3k$, $\sqrt{k_1} + \sqrt{k_2} + \sqrt{k_3} \leq \sqrt{3k}$. Substituting in \eqref{eq:T2bnd},
\begin{equation}
\Pr\big[T_2 \leq -2\sqrt{3kt} - 3t \big] \leq 6e^{-t}. \label{eq:T2Bound}
\end{equation}
Applying the union bound to \eqref{eq:T1Bound}, \eqref{eq:T2Bound} gives the result in \eqref{eq:lemmaFwd} with $C = 4(1 + \sqrt{3}).$
\end{proof}

\end{document}